\documentclass[journal]{IEEEtran}
\IEEEoverridecommandlockouts
\usepackage{cite}
\usepackage{amsmath,amssymb,amsfonts}
\usepackage{algorithmic}
\usepackage{graphicx}
\usepackage{textcomp}
\usepackage{xcolor}
\usepackage[T1]{fontenc}
\usepackage{color}
\usepackage{caption}
\usepackage{enumerate}
\usepackage{graphicx} 
\usepackage{psfrag}
\usepackage{subfigure}
\usepackage{pdfpages}
\usepackage{multirow}
\usepackage{threeparttable}
\usepackage{array}
\usepackage{dsfont}
\usepackage{subeqnarray}
\usepackage{cases}
\usepackage{makecell}
\usepackage{algorithmic}
\usepackage{url}
\usepackage{stfloats}
\usepackage{footnote}

\allowdisplaybreaks[4]
\begin{document}

\title{Improved Receivers for Optical Wireless OFDM: An Information Theoretic Perspective}

\author{\IEEEauthorblockN{Xiaozhen Liu, ~\IEEEmembership{Student Member,~IEEE},
Jing~Zhou,~\IEEEmembership{Member,~IEEE}, Nuo Huang,~\IEEEmembership{Member,~IEEE}, and~Wenyi~Zhang,~\IEEEmembership{Senior~Member,~IEEE}
}

\thanks
{This work was supported in part by the National Natural Science Foundation of China under Grant 62101526, in part by the China Postdoctoral Science Foundation under Grant 2019M662197, in part by the Fundamental Research Funds for the Central Universities under Grant KY2100000118, and in part by the start-up funds from Shaoxing University. This paper was presented in part at Workshop on Optical Wireless Communication (OWC) in VTC 2021 Fall \cite{VTC} (\emph{Xiaozhen Liu and Jing Zhou are co-first authors}).

Xiaozhen Liu, Nuo Huang, and Wenyi Zhang are with CAS Key Laboratory of Wireless-Optical Communications, University of Science and Technology of China, Hefei, China (e-mail: lxz132@mail.ustc.edu.cn; huangnuo@ustc.edu.cn; wenyizha@ustc.edu.cn).

Jing Zhou is with the Department of Computer Science and Engineering, Shaoxing University, Shaoxing, China. He was with CAS Key Laboratory of Wireless-Optical Communications, University of Science and Technology of China, Hefei, China (e-mail: jzhou@usx.edu.cn).}
}

\maketitle

\begin{abstract}
We consider performance enhancement of asymmetrically-clipped optical orthogonal frequency division multiplexing (ACO-OFDM) and related optical OFDM schemes, which are variations of OFDM in intensity-modulated optical wireless communications.
Unlike most existing studies on specific designs of improved receivers, this paper investigates information theoretic limits of all possible receivers.
For independent and identically distributed (IID) complex Gaussian inputs, we obtain an exact characterization of information rate of ACO-OFDM with improved receivers for all SNRs.
It is proved that the high-SNR gain of improved receivers asymptotically achieve $1/4$ bits per channel use, which is equivalent to $3$ dB in electrical SNR or $1.5$ dB in optical SNR;
as the SNR decreases, the maximum achievable SNR gain of improved receivers decreases monotonically to a non-zero low-SNR limit, corresponding to an information rate gain of $36.3\%$.
For practically used constellations, we derive an upper bound on the gain of improved receivers.
Numerical results demonstrate that the upper bound can be approached to within $1$ dB in optical SNR by combining existing improved receivers and coded modulation.
We also show that our information theoretic analyses can be extended to Flip-OFDM and PAM-DMT.
Our results imply that, for the considered schemes, improved receivers may reduce the gap to channel capacity significantly at low-to-moderate SNR.
\end{abstract}


\section{Introduction}
Over the past decades orthogonal frequency division multiplexing (OFDM) has played a vital role in radio frequency (RF) communications.
Recently, much attention has been paid to the design and performance of OFDM in intensity modulation and direct detection (IM/DD) based optical wireless communications, such as indoor visible light communications (VLC) \cite{Lowery}. 
The challenge therein stems from the fact that, the information-carrying optical intensity (defined as the optical power per unit area \cite{FP}) is real-valued and nonnegative (unipolar), while the baseband signal of OFDM is in general complex-valued.
By introducing additional frequency-domain constraints (e.g., Hermitian symmetry, null subcarriers)
combined with certain time-domain processing (e.g., adding direct-current bias, clipping),
a variety of unipolar OFDM schemes have been proposed, among which direct-current biased optical OFDM (DCO-OFDM) and asymmetrically clipped optical OFDM (ACO-OFDM) are two extensively studied benchmark schemes \cite{Lowery,B94,AL06}.
However, such additional constraints and processing for unipolar design may limit the performance; see, e.g., \cite{DA13,zhou2017capacity}.

This paper considers performance enhancement of ACO-OFDM and two closely related optical wireless OFDM schemes, namely Flip-OFDM \cite{fernando2012flip} and pulse-amplitude-modulated discrete multitone modulation (PAM-DMT) \cite{PAM-DMT}.
It is well known that the three schemes have essentially the same performance.
At high signal-to-noise-ratio (SNR), the performance of ACO-OFDM is limited because its even-indexed subcarriers are not modulated.
This problem can be solved by layered schemes \cite{CKE,EL14,W15,TVH15},
which outperform DCO-OFDM and approach the high-SNR capacity of the optical intensity channel \cite{zhou2017capacity,Kschischang2004capacitySPUB,Moser2005,Lapidoth_Moser_Wigger09,farid2010capacity}.
However, at low SNR adding layers is no longer beneficial and the information rate of ACO-OFDM derived in \cite{li2007channel} is far away from the channel capacity \cite{zhou2017capacity}.

Since the invention of ACO-OFDM \cite{AL06}, most studies including the information rate derivation in \cite{li2007channel} assumed a simple receiver that utilizes only the odd subcarriers (referred to as conventional receiver in the sequel).
It has been well known that the conventional receiver is suboptimal because the discarded data at the even subcarriers contain useful information \cite{CKE09,WA09}.
In light of this fact, various improved receivers for ACO-OFDM have been proposed;
see, e.g., \cite{CKE09,WA09,Mean_square,katz2012new,chen2014improved,asadzadeh2011receiver,dang2013novel,dang2018soft} (see also \cite{fernando2012flip,huang2014receiver,huang2015iterative,XZDW15} for related unipolar schemes),
which provided (mostly uncoded) error performance results for different modulation schemes with specific parameter settings.
In \cite{Mean_square,fernando2012flip}, it was noted that at high SNR certain improved receivers achieve an asymptotic gain of $3$ dB in electrical SNR.
But the results therein were obtained by receiver SNR analysis with approximations, rather than direct evaluation of error probability or achievable rate.
It is still not clear whether the $3$ dB gain is the ultimate performance limit.
More importantly, at finite SNR the maximum achievable gain of improved receivers is unknown.
Therefore, it is essential to establish fundamental limits (e.g., the maximum achievable information rate) of improved receivers for ACO-OFDM and related schemes.

In this paper, unlike most existing studies on specific designs of improved receivers, we investigate performance limits of improved receivers from an information theoretic perspective.
In Section II, after introducing ACO-OFDM in the optical intensity channel, we provide comprehensive uncoded error performance comparisons for existing improved receivers, and subsequently raise a series of questions to be answered, which motivate our work.
In Section III, assuming independent and identically distributed (IID) complex Gaussian inputs,
we prove that the maximum achievable information rate of ACO-OFDM is the sum of two terms: (i) the information rate of the conventional receiver, and (ii) a single-letter conditional mutual information characterizing the maximum achievable gain of improved receivers in information rate.\footnote{In information theory, a single-letter characterization of channel capacity or information rate is a mutual information expression that depends on a single channel input variable and a single channel output variable.
In contrast to multi-letter characterizations, the single-letter characterization is easy to compute, and it also sheds light on practical coding techniques. See [\ref{NIT}, Chapter 4.3] for more discussions.}
By asymptotic analysis, we prove that the high-SNR limit of the maximum achievable gain of improved receivers is indeed $3$ dB in electrical SNR, or $1.5$ dB in optical SNR, corresponding to an information rate gain of exactly $1/4$ bits per channel use (c.u.).
We also show that the low-SNR limit of the maximum achievable gain of improved receivers is approximately $0.67$ dB in optical SNR, or $1.35$ dB in electrical SNR.
Numerical evaluations of finite-SNR gains demonstrate that improved ACO-OFDM receivers may reduce the gap to capacity significantly, especially at low-to-moderate SNR.
The residual gap to the capacity of the optical intensity channel is analyzed in Section IV. 
Section V-A extends the information theoretic analyses to practically used constellations such as quadrature amplitude modulation (QAM) and phase-shift keying (PSK).
Section V-B discusses the performance of the so-called genie receiver, which was employed as a performance upper bound of improved receivers in the literature.
Section V-C extends our information theoretic analyses to Flip-OFDM and PAM-DMT.
In Section VI, numerical results on error performance of ACO-OFDM with improved receivers and low-density parity check (LDPC) coded modulation are provided,
which demonstrate that the bounds given in Section V-A can be approached to within 1 dB in optical SNR.
Finally, Section VI provides concluding remarks on our contribution and future study.

\emph{Notation}:
We use $I(X;Y)$ to denote the mutual information between $X$ and $Y$, and $h(\cdot)$ to denote the differential entropy.
We use $\mathcal I$ to denote the information rate of ACO-OFDM per c.u.
Notations $\overline{\mathcal I}$ and $\underline{\mathcal I}$ stand for upper and lower bounds of $\mathcal I$, respectively.
Notations like $\mathcal I_\textrm D$ denotes an information rate under a given input distribution $\textrm D$.
Channel capacity per c.u. is denoted by $\mathcal C$.
Frequency domain quantities are denoted by tilde symbols such as $\tilde X$.
The sign of a real number $a$ is denoted by $\mathrm{sgn}(a)$, where we define $\mathrm{sgn}(0)=1$.
If the components of $\mathbf A=[A_0,...,A_N]$ are IID random variables, then we use $A$ to denote a random variable with the same distribution as $A_i$.
We denote the high-SNR asymptotic relationship
\begin{equation}
\lim\limits_{\mathsf{SNR}\to\infty}\left[A(\mathsf{SNR})-B(\mathsf{SNR})\right]=0
\end{equation}
by $A(\mathsf{SNR})\doteq B(\mathsf{SNR})$, or simply $A\doteq B$ if there is no danger of confusion. 

\section{ACO-OFDM and Improved Receivers}

\subsection{Channel Model}
Consider the discrete-time optical intensity channel \cite{Lapidoth_Moser_Wigger09,farid2010capacity,li2007channel,Chaaban}
\begin{equation}
\label{DTOIC}
R_i=g\cdot S_i+W_i,\;\; S_i\ge 0,\;\; g> 0,
\end{equation}
where the channel input $S_i$ is the instantaneous optical intensity which carries information, the channel output $R_i$ is the photocurrent at the receiver (a photodetector), $g$ is a time-invariant channel gain (including path losses, detector responsivity (in Ampere per Watt), etc.), and $\{W_i\}$ are IID $\mathcal N(0,\sigma^2)$ noise samples.
For a detailed discussion of the model (\ref{DTOIC}), see \cite{B94,Chaaban}.

We define two types of SNRs:
(i) optical SNR with respect to the average optical intensity as $\mathsf {SNR}_\textrm o=\frac{\mathrm E[S]}{\sigma}$ (see \cite{zhou2017capacity,Kschischang2004capacitySPUB,Moser2005,Lapidoth_Moser_Wigger09,farid2010capacity,asadzadeh2011receiver,chaaban16});\footnote{
The Optical SNR here should not be confused with OSNR (see, e.g., \cite{OSNR}), which is commonly used in optically amplified communications systems and causes signal-dependent electrical noise. See [\ref{FP}, Chapter 14.5] for more details.}
(ii) electrical SNR (or receiver SNR) with respect to the power of the photocurrent
at the receiver as $\mathsf {SNR}_\textrm e =\frac{g^2\mathrm E[S^2]}{\sigma^2}$ (see \cite{CKE09,WA09,Mean_square,EL14,chen2014improved,fernando2012flip,TVH15}).
These definitions imply a square-law relationship as $\mathsf {SNR}_\textrm e =\alpha g^2\mathsf {SNR}_\textrm o^2$, where
the factor $\alpha$ cannot be determined unless the distribution of the channel input has been prescribed (an example will be given in Sec. III-A; see equation (\ref{SNRs}) therein).\footnote{In the literature the SNR may also be measured via energy per bit, such as $\mathrm {E_b(opt)}=\mathrm E[S]/b$ and $\mathrm {E_b(elec)}=g^2\mathrm E[S^2]/b$ where $b$ is the number of bits per symbol. There is also a square-law relationship as $\mathrm {E_b(elec)}\propto bg^2\mathrm {E_b^2(opt)}$.}
Therefore, every dB increase of optical power leads to a $2$-dB improvement in electrical SNR.
This is essentially due to the square-law characteristic of direct-detection via a photodiode.
Without loss of generality, we normalize $g$ to unity hereinafter, and
thus the definition of the electrical SNR is simplified to $\mathsf {SNR}_\textrm e =\frac{\mathrm E[S^2]}{\sigma^2}$.

In this paper, we primarily consider an average intensity constraint on the channel input as
\begin{equation}
\label{E}
\mathrm E \left[S\right]\leq {\mathcal E},
\end{equation}
which is equivalent to a constraint on the optical SNR.
But the electrical SNR constraint will also be considered.

\subsection{Principle of ACO-OFDM and Its Conventional Receiver}\label{IIB}

\begin{figure}[t]
\centering
\includegraphics[width=8.5cm]{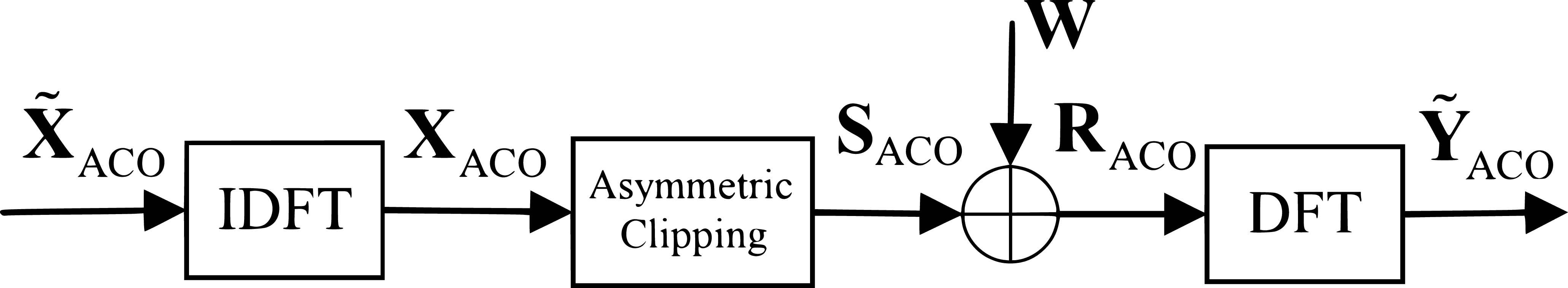}
\caption{Transmission of a symbol block $\tilde {\mathbf{X}}_{\textrm{ACO}}$ by ACO-OFDM.}
\label{ACOfigure}
\end{figure}

ACO-OFDM is a unipolar OFDM scheme for communications over the optical intensity channel \cite{AL06}; see Figure \ref{ACOfigure}.
At an ACO-OFDM transmitter, all even-indexed subcarriers are set to zero, and all odd-indexed ones are modulated.
We assume that there are $N$ subcarriers in total and let $N/4$ be an integer (it can always be satisfied by using a few null subcarriers).
A block of frequency-domain symbols in ACO-OFDM can be written as
\begin{align}
\label{Xtilde}
\tilde {\mathbf{X}}_{\textrm{ACO}}=
\left[ {0,{{\tilde X}_1},0,{{\tilde X}_3},...,{{\tilde X}_{\frac{N}{2} - 1}},0,\tilde X_{\frac{N}{2} - 1}^*,...,\tilde X_3^*,0,\tilde X_1^*} \right],
\end{align}
where the inputs $\{\tilde X_n,\mspace{4mu}n=1,3,...,\frac{N}{2}-1\}$ are drawn independently from an alphabet $\mathcal X$.
By applying the inverse discrete Fourier transform (IDFT), we obtain a block $\mathbf X_{\textrm{ACO}}=[X_0,...,X_{N-1}]$  
which is real-valued since $\tilde {\mathbf{X}}_{\textrm{ACO}}$ is Hermitian.
Since $\mathbf X_{\textrm{ACO}}$ is a superposition of the odd subcarriers, its components satisfy
\begin{equation}
\label{oddsymmetry}
X_i=-X_{i + \frac{N}{2}}, \mspace{4mu}i=  0,...,N/2 - 1.
\end{equation}
Thus we write
\begin{equation}
\mathbf X_{\textrm{ACO}}=\left[\mathbf X, -\mathbf X\right], 
\end{equation}
where
${\mathbf{X}} = [X_0,X_1,...,X_{N/2 - 1}]$. 

By asymmetrically clipping the block $\mathbf X_{\textrm{ACO}}$, we obtain a block of unipolar time samples $\mathbf S_\textrm{ACO} = \left[ {{S_0},{S_1},...,{S_{N - 1}}} \right]$ as the input of the optical intensity channel,
where
\begin{equation}
\label{S}
S_i= \max ({X_i},0)=\frac{1}{2}X_i+\frac{1}{2}|X_i|.
\end{equation}
The block $\mathbf S_\textrm{ACO}$ contains at least $N/2$ zero components corresponding to the positions with $X_i\leq 0$.
We denote the discrete Fourier transform (DFT) of $\mathbf S_\textrm{ACO}$ by $\tilde{\mathbf S}_\textrm{ACO}=[\tilde S_0,...,\tilde S_{N-1}]$.
Note that (\ref{S}) is a decomposition of $\{S_i\}$ such that $\{X_i/2\}$ consists of the odd subcarriers and $\{|X_i|/2\}$ consists of the even ones (generated by asymmetric clipping), if and only if the inputs on even subcarriers are set to zero as (\ref{Xtilde}).
Thus the odd-indexed components of $\tilde{\mathbf S}_\textrm{ACO}$ satisfy
\begin{equation}
\label{Sodd}
\tilde S_n=\frac{1}{2}\tilde X_n, \mspace{4mu}n=1,3,...,N-1.
\end{equation}

At the receiver front-end, a block of channel output is
\begin{equation}
\label{YSW}
\mathbf{R}_\textrm{ACO} = \mathbf{S}_\textrm{ACO}+\mathbf{W},
\end{equation}
where $\mathbf{R}_\textrm{ACO}={[{R_0},{R_1},...,{R_{N-1}}]}$ and
$\mathbf{W}=[{W_0},{W_1},...,{W_{N-1}}]$. 
The receiver performs the DFT and obtains $\tilde{\mathbf Y}_\textrm {ACO}=[{\tilde Y}_0,...,{\tilde Y}_{N-1}]$.
Based on the property (\ref{Sodd}),
the \emph{conventional receiver} of ACO-OFDM discards the even subcarriers in $\tilde{\mathbf Y}_\textrm {ACO}$ and observes the odd ones, thereby obtaining $N/4$ \emph{independent} parallel channels in the frequency domain:
\begin{equation}
\label{FDC}
\tilde Y_n=\frac{1}{2}\tilde X_n +\tilde W_n, \mspace{4mu}n=1,3,...,N/2-1,
\end{equation}
where $\tilde W_n\sim\mathcal N(0,\sigma^2)$.

Combining (\ref{oddsymmetry}), (\ref{S}), and (\ref{YSW}), for $n=0,...,N/2-1$ we have
\begin{subequations}
\begin{align}
&R_i-R_{i + \frac{N}{2}}=X_i+W_i-W_{i + \frac{N}{2}},\\
&R_i+R_{i + \frac{N}{2}}=|X_i|+W_i+W_{i + \frac{N}{2}}.
\end{align}
\end{subequations}
We let
$Y_{1,i}=R_i-R_{i + \frac{N}{2}}$, $Y_{2,i}=R_i+R_{i + \frac{N}{2}}$,
$Z_{1,i}=W_i-W_{i + \frac{N}{2}}$, and $Z_{2,i}=W_i+W_{i + \frac{N}{2}}$,
obtaining an equivalent expression of (\ref{YSW}) as
\begin{subequations}\label{Y1Y2}
\begin{align}
&{{\mathbf{Y}}_1} = {\bf{X}} + {{\bf{Z}}_1}, \label{Y1}\\
&{{\mathbf{Y}}_2} = |{\bf{X}}| + {{\bf{Z}}_2}\label{Y2},
\end{align}
\end{subequations}
where ${\mathbf{Y}}_j=[Y_{j,0},...,Y_{j,\frac{N}{2}-1}]$ and ${\mathbf{Z}}_j=[Z_{j,0},...,Z_{j,\frac{N}{2}}-1]$, in which $j\in\{1,2\}$.
The components of ${{\bf{Z}}_1}$ and ${{\bf{Z}}_2}$ are IID Gaussian with variance
\begin{equation}\label{sigmaz}
\sigma_z^2=2\sigma^2,
\end{equation}
since they are jointly Gaussian satisfying
\begin{align}
\mathrm E[Z_{j,i} Z_{k,i'}]=
\begin{cases}
2\sigma^2, &i=i',\mspace{4mu}j=k,\\
0, &\textrm{otherwise},\\
\end{cases}
\end{align}
where $j,k\in\{1,2\}$.

We can split $\tilde{\mathbf Y}_\textrm {ACO}$ into
\begin{subequations}
\begin{align}
&{\tilde {\mathbf{Y}}_{\textrm {odd}}} = [0, {{{\tilde Y}_1},0,{{\tilde Y}_3},...0,{{\tilde Y}_{N - 1}}} ],\\ 
&{\tilde {\mathbf{Y}}_{\textrm {even}}} = [ {{{\tilde Y}_0},0,{{\tilde Y}_2},0,...,{{\tilde Y}_{N - 2}}},0 ]. 
\end{align}
\end{subequations}
It is straightforward to show that $[\mathbf Y_1, -\mathbf Y_1]$ is the IDFT of $2\tilde {\mathbf{Y}}_{\textrm {odd}}$, and $[\mathbf Y_2, \mathbf Y_2]$ is the IDFT of $2\tilde {\mathbf{Y}}_{\textrm {even}}$.
This is because the operation $R_i-R_{i + \frac{N}{2}}$ cancels the even subcarriers, while the operation $R_i+R_{i + \frac{N}{2}}$ cancels the odd subcarriers.
Since the conventional receiver observes only $\tilde {\mathbf{Y}}_{\textrm {odd}}$ rather than $\tilde{\mathbf Y}_\textrm {ACO}$,
its performance should be equivalent to a receiver that discards (\ref{Y2}) and observes only (\ref{Y1}).

\subsection{Improved Receivers for ACO-OFDM: Uncoded Performance and Motivation of Our Work}

An \emph{improved receiver} for ACO-OFDM seeks to improve the performance of the conventional receiver (determined by the equivalent channel (\ref{FDC})) by utilizing the structure of the ACO-OFDM signal described above.
Table \ref{tabel111} shows several kinds of improved receivers for ACO-OFDM and related schemes in the literature.
Some improved receivers were proposed for Flip-OFDM, but they can also be utilized in ACO-OFDM through simple modification (see Section \ref{related}).
The first three kinds of receivers are relatively simple since they utilize only time domain signal structures.
The others require more computation costs, e.g., fast Fourier transform/inverse fast Fourier transform (FFT/IFFT) operations, iterations, etc.

\begin{figure}
\includegraphics[scale=0.59]{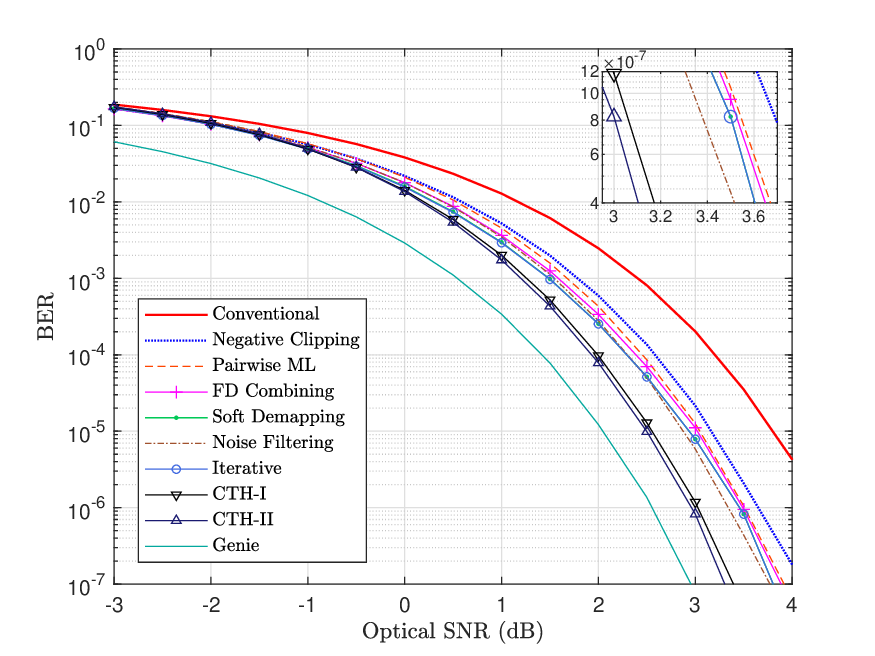}
\caption{BER performance of conventional and improved ACO-OFDM receivers (4-QAM).
}
\label{figure4M4gezhongReceiver}
\end{figure}

\begin{figure}
\centering
\includegraphics[scale=0.59]{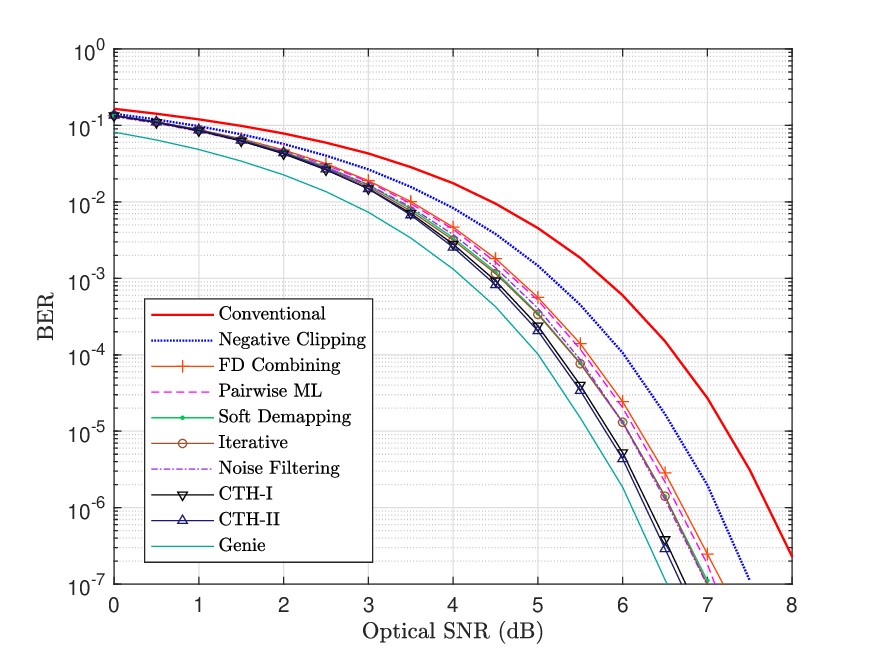}
\caption{BER performance of conventional and improved ACO-OFDM receivers (16-QAM).
}
\label{figure4M16gezhongReceiver}
\end{figure}

\begin{figure}
\centering
\includegraphics[scale=0.59]{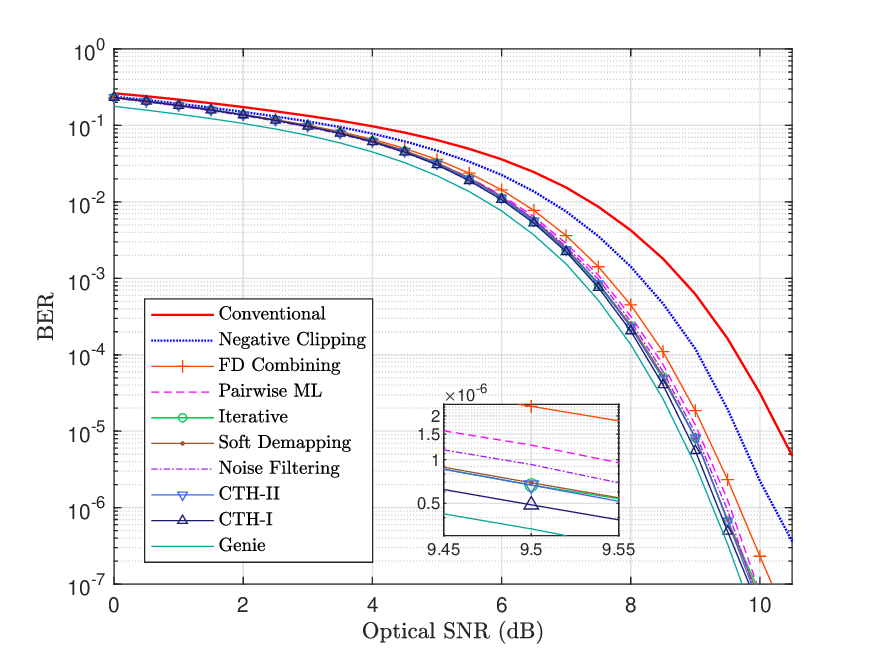}
\caption{BER performance of conventional and improved ACO-OFDM receivers (64-QAM).}
\label{figure5M64gezhongReceiver}
\end{figure}

In the references in Table \ref{tabel111}, only partial comparisons of their performance can be found, and most of the results consider uncoded BER performance with 4-QAM or 16-QAM constellation.
In Figures \ref{figure4M4gezhongReceiver}--\ref{figure5M64gezhongReceiver}, we provide comprehensive uncoded BER comparisons in the optical intensity channel with 4-QAM, 16-QAM, and 64-QAM constellations, respectively.
The performance of the so-called \emph{genie receiver} \cite{asadzadeh2011receiver} with negative clipping is also shown as a BER lower bound.
While a practical receiver cannot guarantee error-free localization of zero components in $[S_0,...,S_{N-1}]$ by observing $\mathbf R$ at a finite SNR,
the genie receiver perfectly knows the locations of $N/2$ zero components therein (or equivalently, it perfectly knows the sign of $\mathbf X$; see (\ref{S})).
The numerical results show that, at high SNR, the maximum SNR gain of improved receivers over the conventional receiver approaches $1.5$ dB in optical SNR.
As the constellation size increases, the performances of several improved receivers get closer, and their gaps to the performance of the genie receiver decrease.

According to the above numerical results, we have the following natural questions with discussions.
\begin{itemize}

\item Although many improved receivers with different approaches have been proposed, there appears to be an inherent performance limit for them.
Specifically, at high SNR, the numerical results on improved receivers show a gain of about $1.5$ dB in optical SNR.
A related result is the achievability of the $3$ dB asymptotic gain (i.e., a factor of two) in electrical SNR, which has been observed in, e.g., \cite{CKE09,Mean_square,katz2012new,fernando2012flip,dang2018soft}.\footnote{In fact, the $3$ dB asymptotic gain can be achieved by simple improved receivers (e.g. that in \cite{CKE09}) which is clearly suboptimal at finite SNR.}
Without considering the complexity, can we further improve the performance?
In other words, has the fundamental limit of improved receivers been approached by existing ones?

\item
In uncoded performance studies as Figures 2-4, only a few transmission rates (determined by the size of the constellations) can be considered.
Can we establish a complete characterization of the ultimate performance of improved receivers for all rates/SNRs?
How large is the residual gap to the capacity of the optical intensity channel?

\item
For the genie receiver,
a gain of about $1.5$ dB (up to $2$ dB at low SNR) with respect to the performance of the conventional receiver in optical SNR was observed in \cite{asadzadeh2011receiver},
while a gain of $3$ dB (at all SNRs) in electrical SNR was observed in \cite{chen2014improved} (see also \cite{dang2013novel,huang2015iterative,huang2014receiver} for more results).
The gains were only intuitively interpreted (e.g., by claiming that half of the noise power can be removed).
Can we give an exact characterization of the performance of the genie receiver, and give a clear interpretation of the above observations?
\end{itemize}

\begin{table}[t]
\centering
\captionof{table}{Improved Receivers for ACO-OFDM and Related Schemes}
\label{tabel111}
\scalebox{1.1}{
\begin{tabular}{|c|l|}
\hline
\multicolumn{1}{|c|}{Name} & \multicolumn{1}{c|}{References}
\\ \hline
Negative Clipping  & \cite{WA09}
\\ \hline
\makecell[cc]{Pairwise ML\\(With Clipping)} & \cite{asadzadeh2011receiver}
\\ \hline
Noise Filtering & \cite{fernando2012flip}
\\ \hline
\makecell[cc]{Frequency-Domain\\(FD) Combining}& \begin{tabular}[c]{@{}l@{}}\cite{CKE09},\cite{katz2012new},\cite{Mean_square}
\end{tabular}
\\ \hline
\makecell[cc]{Iterative Receiver\\(MMSE/ZF)}  & \begin{tabular}[c]{@{}l@{}}\cite{dang2013novel}, \cite{huang2015iterative}\end{tabular}
\\ \hline
Soft Demapping & \cite{dang2018soft}
\\ \hline
\makecell[cc]{Chen-Tsonev-Haas I\\(CTH-I) and CTH-II} & \cite{chen2014improved}
\\ \hline
\end{tabular}}
\end{table}

In view of the lack of theoretical foundation for this topic as discussed above,
the focus of the remaining part of this paper is to address these raised questions using information theoretic methods.
Specifically, we will characterize information theoretic limits of ACO-OFDM with improved receivers by Propositions 1 and 4;
large-SNR asymptotics of the information theoretic limits will be given in Proposition 2;
an analyze of the residual gap to capacity will be given in Table II;
the performance of the genie receiver will be characterized by Proposition 5 and related discussions.

\section{Information Theoretic Limits of ACO-OFDM}\label{SecIII}

\subsection{Preliminaries}
In our information theoretic studies, we characterize the performance of ACO-OFDM by the \emph{information rate}, which is given by the mutual information between the channel input block and output block, normalized by the number of channel uses $N$:
\begin{equation}
\label{IACO}
\mathcal I=\frac{1}{N}I(\tilde{\mathbf X}_\textrm{ACO};\mathbf R_\textrm{ACO}).
\end{equation}
According to the channel coding theorem \cite{Gallager},
this information rate is the maximum coded transmission rate of ACO-OFDM with vanishing error probability as the code length grows without bound.
But it is not necessarily achievable if a specific suboptimal receiver is used.
In fact, the information rate $\mathcal I$ is achievable with the maximum likelihood (ML) receiver, and it is also the maximum achievable information rate of all possible improved receivers.\footnote{Determining the maximum achievable information rate with a specific suboptimal receiver belongs to the topic of \emph{mismatched capacity} \cite{mismatch}, which is beyond the scope of this paper.}
Unfortunately, the complexity of the ML receiver of ACO-OFDM grows exponentially in $N$, because each component of the received block $\mathbf R_\textrm{ACO}$ (or $\tilde {\mathbf Y}_{\textrm{ACO}}$) depends on all $N/4$ inputs.
The task of improved receivers is achieving a better tradeoff between performance and complexity, and even approaching the performance of the ML receiver.

The \emph{capacity} of the optical intensity channel (\ref{DTOIC}) is the maximum achievable information rate over it:
\begin{align} \label{C}
\mathcal C=\max\limits_{p_S}I(S;R),
\end{align}
where $p_S(s)$ satisfies certain constraint, e.g., $\mathrm E[S]\leq\mathcal E$ or $\mathrm E[S]/\sigma=\mathsf {SNR}_\mathrm{o}$.
In contrast to (\ref{IACO}), the capacity (\ref{C}) has a single-letter characterization since the channel (\ref{DTOIC}) is memoryless.
Although no closed-form expression for (\ref{C}) is known, tight bounds can be found in \cite{Kschischang2004capacitySPUB,Moser2005,Lapidoth_Moser_Wigger09,farid2010capacity,chaaban16},
and the high-SNR capacity satisfies \cite{Lapidoth_Moser_Wigger09}
\begin{equation}
\label{ChighSNR}
\mathcal C\doteq\frac{1}{2}\log\frac{e\mathcal E^2}{2\pi\sigma^2}.
\end{equation}
Clearly we have $\mathcal I\leq \mathcal C$, and the gap between them is of interest to us.

In this section, we let the input alphabet be $\mathcal X=\mathbb R$ and
assume that the inputs
${\tilde {{X}}_1}$, ${\tilde {{X}}_3},...,{\tilde {{X}}_{N/2-1}}$ are IID ${\cal CN}(0,\tilde\sigma_x^2)$.
Then $\mathbf X$ is jointly Gaussian with uncorrelated components, i.e.,
$\textrm E\left[X_i X_{i^\prime}\right]=0$, $\forall i\neq i^\prime$.
Consequently, the components of $\mathbf X$ are IID ${\cal N}(0,\sigma_x^2)$,
where $\sigma_x^2=\tilde\sigma_x^2/2$ (we let the DFT/IDFT be normalized so that $\|\tilde{\mathbf X}_\textrm{ACO}\|^2=\|\mathbf X_\textrm{ACO}\|^2$).
In this setting, first, (\ref{Y1}) and (\ref{Y2}) are memoryless, so that
\begin{align}
I(\mathbf {X};\mathbf{Y}_1)             &=\frac{N}{2} I(X;Y_1),\label{memoryless1}\\
I(\mathbf {X};\mathbf{Y}_1,\mathbf{Y}_2)&=\frac{N}{2} I(X;Y_1,Y_2),\label{memoryless2}
\end{align}
where ${{{X}}}\sim {\cal N}(0,\sigma_x^2)$, and $Y_j$ is a random variable with the same distribution as a component of $\mathbf Y_j$, $j\in\{1,2\}$.
Second, for a given time index, the ACO-OFDM signal obeys a clipped Gaussian distribution with a probability density function (PDF) as
\begin{equation}
\label{S_f}
{f_{S_{\textrm{clip}}}}(s) = \frac{{\rm{1}}}{{\rm{2}}}\delta \left( s \right) + \frac{{u \left( s \right)}}{{  \sigma _x }}\phi \left( {\frac{{  {s}}}{{\sigma _x}}} \right),
\end{equation}
where $S_{\textrm{clip}}$ stands for a clipped Gaussian variable, $\delta \left( s \right)$ is the Dirac delta function, ${u \left( s \right)}$ is the unit step function, and
$\phi(a)=(2\pi)^{-\frac{1}{2}}e^{-\frac{a^2}{2}}$
is the PDF of the standard normal distribution.
However, it should be noted that the components of $\mathbf S_\textrm{ACO}$ are not IID.

According to (\ref{S_f}) and the fact $\tilde{\sigma}_x^2=2\sigma_x^2$, we have
\begin{align}
&\mathrm E[S_{\textrm{clip}}]=\frac{\sigma_x}{\sqrt{2\pi}}=\frac{\tilde{\sigma}_x}{2\sqrt{\pi}},\label{EE}\\
&\mathrm E[S_{\textrm{clip}}^2]=\frac{\sigma_x^2}{2}=\frac{\tilde{\sigma}_x^2}{4},\label{ES}
\end{align}
leading to a simple relationship between the two SNRs:
\begin{align}
\label{SNRs}
\mathsf {SNR}_\textrm e=\pi\mathsf{SNR}_\textrm o ^2.
\end{align}
In this case, a $3$ dB gain in electrical SNR is equivalent to a $1.5$ dB gain in optical SNR.
We use $\mathcal I(\mathsf {SNR})$ to denote the mutual information per c.u. achieved at a given SNR,
where the argument $\mathsf {SNR}$ can be either $\mathsf {SNR}_\textrm e$ or $\mathsf{SNR}_\textrm o$, as specified by the context.

\subsection{Information Rate With Conventional Receiver}
Based on the equivalent channel (\ref{FDC}), the information rate of ACO-OFDM with conventional receiver is given by
\begin{align}\label{I_}
\underline {\mathcal I}=\frac{1}{{N}}I(\mathbf {\tilde X}_\textrm {ACO};\mathbf{\tilde Y}_\textrm {odd})
=\frac{1}{{4}}I\left(\tilde X;\frac{1}{2}\tilde X+\tilde W\right),
\end{align}
where $\tilde W \sim\mathcal N(0,\sigma^2)$, and the second equality follows because IID inputs leads to $N/4$ independent frequency-domain subchannels as (\ref{FDC}).
The result under IID complex Gaussian inputs, denoted by $\underline {\mathcal I}_\textrm{G}$, has been given in \cite{li2007channel} as
\begin{align}
\label{piE}
\underline {\mathcal I}_\textrm{G}=\frac{1}{4}\log\left(1+\frac{\pi\mathcal E^2}{\sigma^2}\right),
\end{align}
which also follows from (\ref{I_}) and the relationship (\ref{EE}). 

We note that
\begin{subequations}
\begin{align}
I(\mathbf {\tilde X}_\textrm {ACO};\mathbf{\tilde Y}_\textrm {odd})
&=I(\mathbf { X}_\textrm {ACO};\mathbf{ Y}_1)\label{TR1}\\
&=I(\mathbf { X};\mathbf{ Y}_1)\label{TR2}\\
&=\frac{N}{2}I( X;Y_1),\label{TR3}
\end{align}
\end{subequations}
where (\ref{TR1}) and (\ref{TR2}) follow from the facts that $\mathbf {\tilde X}_\textrm {ACO}\to\mathbf { X}_\textrm {ACO}\to\mathbf {X}$ and $\mathbf{\tilde Y}_\textrm {odd}\to\mathbf{ Y}_1$ are one-to-one mappings, and (\ref{TR3}) follows from (\ref{memoryless1}).
Therefore, under IID complex Gaussian inputs, the information rate of ACO-OFDM with conventional receiver (\ref{piE}) can also be expressed by
\begin{align}
\label{underline222}
\underline {\mathcal I}_\textrm{G}=\frac{1}{2}I(X; Y_1).
\end{align}

\subsection{Information Rate With Improved Receivers}
With improved receivers, the maximum achievable information rate of ACO-OFDM is characterized by (\ref{IACO}), which is difficult to evaluate directly.
Utilizing the structure of the ACO-OFDM signal,
we establish the following result which shows that, under IID complex Gaussian inputs,
the information rate (\ref{IACO}) is the sum of (i) the mutual information $\underline{\mathcal I}_\textrm{G}$, which is the information rate of the conventional receiver, as given by (\ref{piE}), and (ii) a conditional mutual information $\Delta$, as given by (\ref{singleletter}) in Proposition 1 below, which is a single-letter characterization corresponding to the maximum achievable gain of improved receivers.

\emph{Proposition 1: Under IID complex Gaussian inputs, the information rate of ACO-OFDM in the optical intensity channel (\ref{DTOIC}) is given by}
\begin{align}\label{III}
\mathcal I_\textrm{G}=\underline{\mathcal I}_\textrm{G}+\Delta, 
\end{align}
\emph{where} $\underline{\mathcal I}_\textrm{G}$ \emph{is the information rate with the conventional receiver given in (\ref{piE}), and}
\begin{align}
\Delta &=\frac{1}{2}I({ X}; Y_2|Y_1), \label{singleletter}
\end{align}
\emph{where} $X\sim\mathcal{N}(0,2\pi \mathcal E^2)$\emph{,} $Y_1=X+Z_1$\emph{, and} $Y_2=|X|+Z_2$\emph{, where} $Z_1$ \emph{and} $Z_2$ \emph{are independent and satisfy} $Z_i\sim \mathcal N(0,2\sigma^2),\; i\in\{1,2\}$.

\begin{IEEEproof}
Assuming IID complex Gaussian inputs in the frequency domain,
we can simplify (\ref{IACO}) as follows:
\begin{subequations}
\begin{align}
\mathcal I_\textrm{G}&=\frac{1}{N}I(\mathbf {\tilde X}_\textrm {ACO};\mathbf{R}_\textrm{ACO})\\
&=\frac{1}{N}I(\mathbf { X}_\textrm {ACO};\mathbf{R}_\textrm{ACO})\label{I1}\\
&=\frac{1}{N}I(\mathbf { X}; \mathbf { Y}_1, \mathbf { Y}_2)\label{I3}\\
&=\frac{1}{2}I(X;Y_1,Y_2)\label{I4}\\
&=\frac{1}{2}I(X;Y_1)+\frac{1}{2}I(X;Y_2|Y_1),\label{I5}
\end{align}
\end{subequations}
where (\ref{I1}) and (\ref{I3}) follow because $\mathbf {\tilde X}_\mathrm {ACO}\to\mathbf { X}_\mathrm {ACO}\to\mathbf {X}$ and $\mathbf R_\textrm{ACO}\to\{\mathbf Y_1,\mathbf Y_2\}$ are one-to-one mappings, and (\ref{I4}) follows from (\ref{memoryless2}).
Combining (\ref{I5}) with (\ref{underline222}) we obtain (\ref{III}).
The proof is completed by noting that, to maximize the information rate, we should let $\sigma _x=\sqrt {2\pi }\mathcal E$ (i.e., let $\mathrm E[S]$ achieve its upper limit $\mathcal E$, see (\ref{EE})).
\end{IEEEproof}

Since
\begin{subequations}
\begin{align}
I({ X}; Y_2|Y_1)&=h(Y_2|Y_1)-h(Y_2|X,Y_1)\\
&=h(Y_2|Y_1)-h(Z_2)\\
&=h(Y_2|Y_1)-\frac{1}{2}\log\left( { 4\pi e\sigma^2}\right),
\end{align}
\end{subequations}
the problem of evaluating information theoretic limits of ACO-OFDM with improved receivers then boils down to evaluating the conditional differential entropy:
\begin{subequations}
\begin{align}\label{condition_entropy}
&h(Y_2| Y_1)=\mathrm E\left[\log\frac{1}{p_{Y_2|Y_1}(y_2|y_1)}\right]\\
&=\int_{-\infty}^{\infty} p_{Y_1}(y_1)\int_{-\infty}^{\infty} p_{Y_2|Y_1}(y_2|y_1)\log\frac{1}{p_{Y_2|Y_1}(y_2|y_1)}dy_2dy_1.
\end{align}
\end{subequations}
\subsection{Evaluation of $h(Y_2|Y_1)$}
\begin{figure}
\centering
\includegraphics[scale=.59]{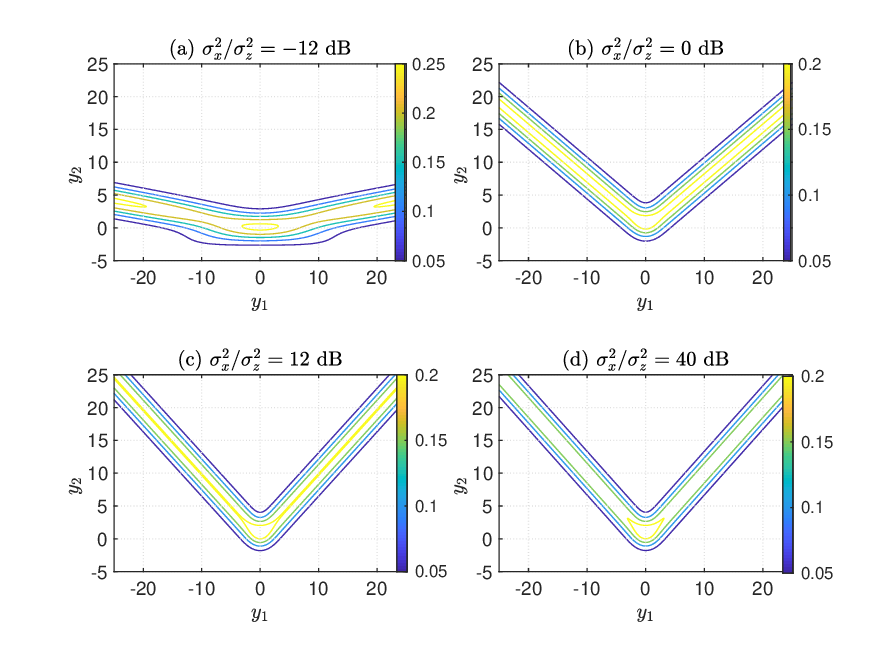}
\caption{The conditional PDF $p_{{Y_2}|{Y_1}}({y_2}|{y_1})$.
}
\label{figure:Conditional_diatribution}
\end{figure}

\begin{figure}
\centering
\includegraphics[scale=.59]{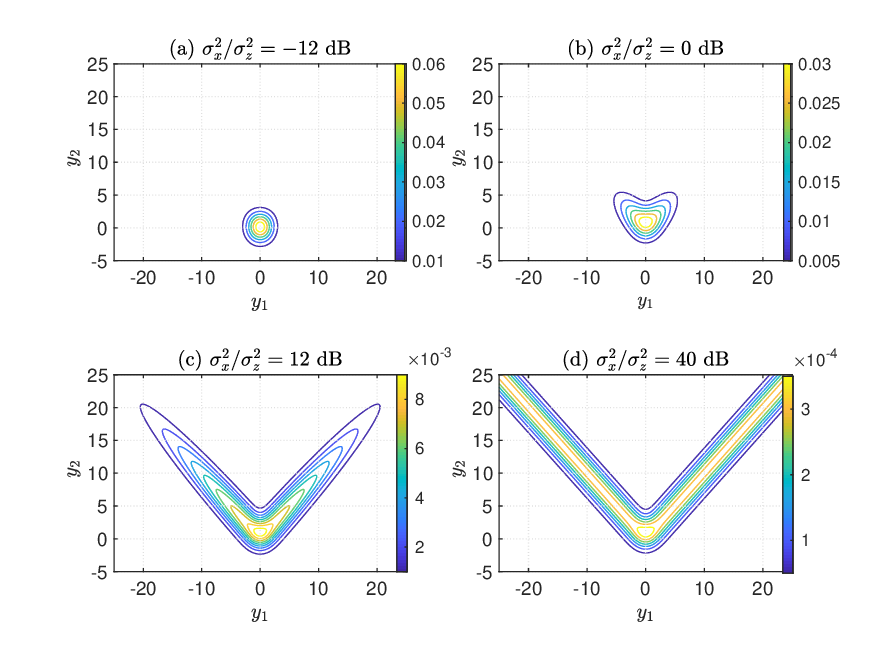}
\caption{The joint PDF $p_{{Y_2},{Y_1}}({y_2},{y_1})$.}
\label{figure:Joint_diatribution}
\end{figure}

To derive the conditional PDF $p_{Y_2|Y_1}(y_2|y_1)$ in (\ref{condition_entropy}),
we begin with
\begin{subequations}
\begin{align}
 {p_{{Y_2}|{Y_1}}}({y_2}|{y_1}) &= \int_{ - \infty }^{ + \infty } {{p_{{Y_2,Z_1}|{Y_1}}}({y_2,z_1}|{y_1})d{z_1}}\\
 &= \int_{ - \infty }^{ + \infty } {{p_{{Z_1}|{Y_1}}}({z_1}|{y_1}){p_{{Y_2}|{Y_1},{Z_1}}}({y_2}|{y_1},{z_1})d{z_1}}, \label{32}
\end{align}
\end{subequations}
where the second conditional PDF in (\ref{32}) is
\begin{eqnarray}
\label{33}
{p_{{Y_2}|{Y_1},{Z_1}}}\left( {{y_2}|{y_1},{z_1}} \right) =\frac{1}{\sigma_z}\phi\left(\frac{y_2-|y_1-z_1|}{\sigma_z}\right).
\end{eqnarray}
The first conditional PDF in (\ref{32}) can be obtained by
\begin{align}
p_{{Z_1}|{Y_1}}({z_1}|{y_1})=\frac{p_{Y_1|Z_1}({y_1}|{z_1})p_{Z_1}(z_1)}{p_{Y_1}(y_1)},
\end{align}
where
\begin{align}
p_{Y_1|Z_1}({y_1}|{z_1})=\frac{1}{\sigma_x}\phi\left(\frac{y_1-z_1}{\sigma_x}\right).
\end{align}
Through some manipulations we obtain
\begin{subequations}\label{35}
\begin{align}
  &{p_{{Z_1}|{Y_1}}}({z_1}|{y_1})\\
  &=\frac{\sqrt{\sigma _x^2 + \sigma _z^2}} {\sigma _x\sigma _z}\phi\left(\frac{\sqrt{\sigma _x^2 + \sigma _z^2}} {\sigma _x\sigma _z}\left(z_1-\frac{\sigma_z^2}{\sigma_x^2+\sigma_z^2}y_1\right)\right).
\end{align}
\end{subequations}
Substituting (\ref{33}) and (\ref{35}) into (\ref{32}) yields
\begin{align}
\label{CPDF}
{p_{{Y_2}|{Y_1}}}&({y_2}|{y_1})=
{f({y_1},{y_2})g( - {y_1},{y_2}) + f( - {y_1},{y_2})g({y_1},{y_2})},
\end{align}
where
\begin{eqnarray}
f({y_1},{y_2}) = \frac{1}{\sigma_f}\phi
\left(\frac{1}{\sigma_f}\left(y_2+\frac{\sigma_x^2}{\sigma_x^2+2\sigma^2}y_1\right)\right)
\end{eqnarray}
with
\begin{align}
\sigma_f=2\sqrt{\frac{\sigma^2(\sigma_x^2+\sigma^2)}{\sigma_x^2+2\sigma^2}},
\end{align}
and
\begin{eqnarray}
g({y_1},{y_2}) = \Phi \left( \frac{\sigma_x(y_2+y_1)}{{2\sigma\sqrt {\sigma_x^2+\sigma^2} }} \right)
\end{eqnarray}
with
$\Phi \left( x \right) = \int_{ - \infty }^x \phi(u)du$.
We plot the conditional PDF $p_{{Y_2}|{Y_1}}({y_2}|{y_1})$ and the joint PDF $p_{{Y_2},{Y_1}}({y_2},{y_1})$ under different SNRs in Figure \ref{figure:Conditional_diatribution} and Figure \ref{figure:Joint_diatribution}, respectively,
where we use contour lines (isodensity lines) to show the two-dimensional PDFs in the $(y_1,y_2)$-plane.
Then the conditional differential entropy $h(Y_2|Y_1)$ given in (\ref{condition_entropy}) can be evaluated based on (\ref{CPDF}) and the fact $Y_1\sim \mathcal N(0,\sigma_x^2+2\sigma^2)$.


\subsection{High-SNR Limit}
The following result shows that the high-SNR limit of the maximum achievable gain of improved ACO-OFDM receivers is $1/4$ bits/c.u. in information rate, or $1.5$ dB in optical SNR, or $3$ dB in electrical SNR.

\emph{Proposition 2: For ACO-OFDM with IID complex Gaussian inputs, at high SNR, the maximum gain of improved receivers satisfies}
\begin{align}\label{Gainbit}
\Delta=\mathcal{I}_\textrm{G} -\underline {\mathcal I}_\textrm{G}\doteq\frac{1}{4}  \mspace{4mu}{\textrm {bits/c.u.}},
\end{align}
\emph{and the maximum achievable information rate with improved receivers satisfies the following equivalent asymptotic expressions:}
\begin{align}
\mathcal{I}_\textrm{G}(\mathsf{SNR}_\textrm o)&\doteq \frac{1}{4}\log\frac{2\pi\mathcal E^2}{\sigma^2},\label{IhighSNR}\\
\mathcal{I}_\textrm{G}(\mathsf {SNR}_\textrm o) &\doteq \underline {\mathcal I}_\textrm{G}(\sqrt{2}\cdot\mathsf {SNR}_\textrm o),\label{GainSNR}\\
\mathcal{I}_\textrm{G}(\mathsf {SNR}_\textrm e)&\doteq\underline {\mathcal I}_\textrm{G}(2\cdot\mathsf {SNR}_\textrm e),\label{GainSNRe}
\end{align}
\emph{where} $\underline {\mathcal I}_\textrm{G}$ \emph{is the information rate with the conventional receiver given by (\ref{piE}).}

\begin{IEEEproof}
According to (\ref{Y1Y2}), we have
\begin{eqnarray}
\label{cases}
{Y_2}  =\left\{ \begin{gathered}
  | Y_1|- {Z_1}  + {Z_2},\;\;\;\;\mspace{4mu} \textrm{if} \; Y_1 > 0, \mspace{4mu}Y_1 > Z_1  \hfill \\
    -| Y_1|+{Z_1}  + {Z_2},\; \;\textrm{if} \; Y_1 > 0, \mspace{4mu}Y_1 \leq Z_1 \hfill \\
  -|{{Y_1}} |- {Z_1} + {Z_2},\;\;\textrm{if} \; Y_1 \leq 0,\mspace{4mu}Y_1 > Z_1 \hfill \\
   |{{Y_1}} |+ {Z_1} + {Z_2},\;\; \;\;\mspace{4mu}\textrm{if} \; Y_1 \leq 0,\mspace{4mu}Y_1 \leq Z_1. \hfill \\
\end{gathered}  \right.
\end{eqnarray}
Since
$\Pr\left(\left| {{Y_1}} \right| > \left| {{Z_1}} \right|\right)\to 1$ as the SNR grows without bound,
the probability of the second and the third cases in (\ref{cases}) tends to zero,
implying that the high-SNR limit of ${p_{{Y_2}|{Y_1}}}({y_2}|{y_1})$ is
\begin{align}
{p^*_{{Y_2}|{Y_1}}}({y_2}|{y_1}) = \frac{1}{{2\sigma }}\phi \left( { \frac{{{{{y_2} - |{y_1}|}}}}{{2\sigma}}} \right),
\end{align}
and consequently the differential entropy $h(Y_2|Y_1)$ converges to that of ${p^*_{{Y_2}|{Y_1}}}({y_2}|{y_1})$ which is $h^*(Y_2|Y_1)=\frac{1}{2}{\log}\left( {8\pi e \sigma^2} \right)=\frac{1}{2}{\log}\left( {4\pi e \sigma _z^2} \right)$.
Therefore
\begin{align}
\mspace{-12mu}\frac{1}{2}I(X;Y_2|Y_1)\doteq\frac{1}{2}(h^*(Y_2| Y_1)-h(Z_2))=\frac{1}{4} \mspace{4mu}{\textrm {bits}},
\end{align}
and we obtain (\ref{Gainbit}).
Finally, the proof is completed by noting that (\ref{GainSNR}) follows from (\ref{piE}) and (\ref{Gainbit}),
and (\ref{GainSNRe}) follows from (\ref{SNRs}) and (\ref{GainSNR}).
\end{IEEEproof}

\subsection{Numerical Results and Discussions}
\begin{figure*}
\centering
\includegraphics[scale=0.8]{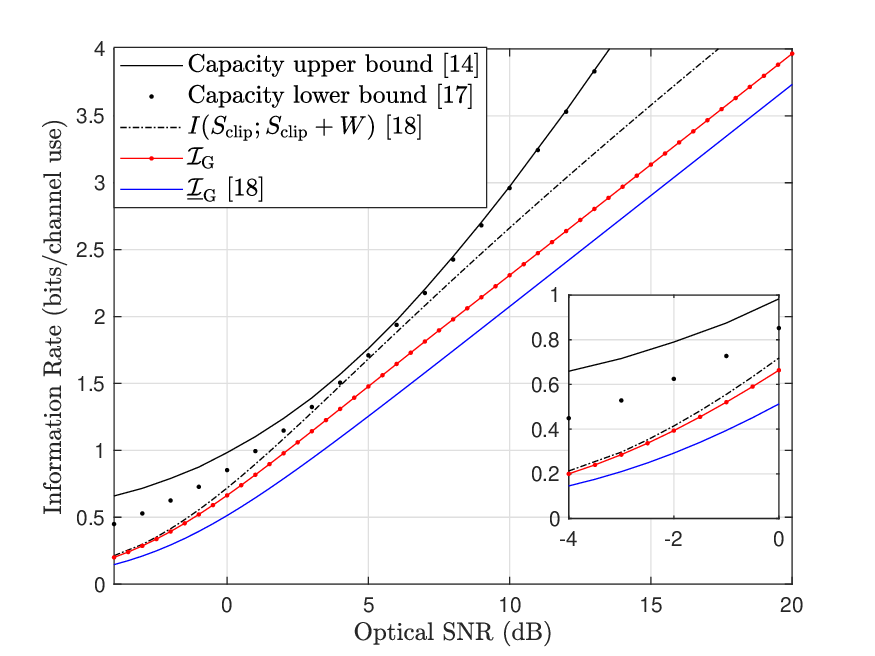}
\caption{Comparison of achievable information rates of ACO-OFDM.}
\label{Ifigure}
\end{figure*}

In Figure \ref{Ifigure}, we show the maximum achievable information rate of ACO-OFDM with improved receivers given in Proposition 1, and compare it with channel capacity bounds and the information rate achieved by the conventional receiver.
At low-to-moderate SNR, the receiver improvement may reduce the gap to capacity significantly.
Specifically, when the optical SNR is $0$ to $5$ dB, the gap to the best-known capacity lower bound can be reduced by more than a half.

\begin{figure}
\centering
\includegraphics[scale=0.59]{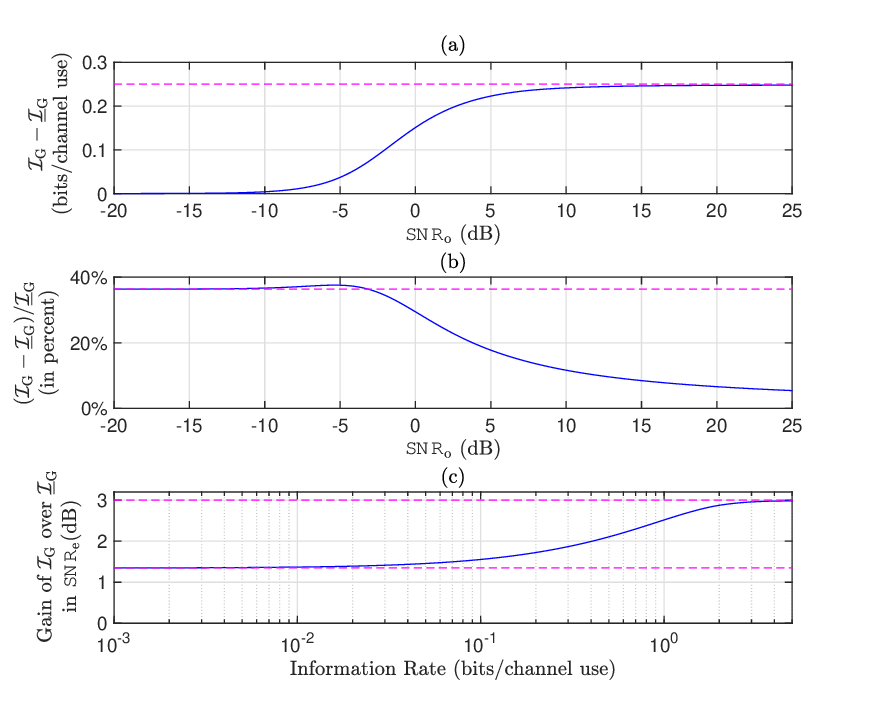}
\caption{Maximum achievable gain of improved ACO-OFDM receivers.}
\label{gain}
\end{figure}

In Figure \ref{gain}, we show the maximum achievable gain of improved ACO-OFDM receivers in different ways.
The $3$ dB high-SNR limit of the gain in electrical SNR is validated, and we further show that the gain decreases monotonically as the SNR decreases, thus validating that $3$ dB is the maximum possible gain for all SNRs.
Interestingly, the relative gain in information rate is larger at low SNR, and it is maximized when the optical SNR is around $-5$ dB.
A further finding is that the low-SNR limit of gain is approximately $0.67$ dB in optical SNR or $1.35$ dB in electrical SNR;
equivalently, in the low-SNR limit the relative gain of information rate is approximately $36.3\%$.

In Figure \ref{figure:Conditional_diatribution}, (a)-(d) show increasing dependencies between $Y_1$ and $Y_2$ as the SNR increases, which imply that more information can be extracted from $Y_2$ at high SNR.
An intuitive interpretation of the high-SNR asymptotic behavior of the gain $\Delta=\frac{1}{2}I(X;Y_2|Y_1)$ is as follows.
Since we already have an observation $Y_1=X+Z_1$ (corresponding to $\underline{\mathcal I}_\mathrm G=\frac{1}{2}I(X;Y_1)$, the information rate with conventional receiver), the conditional mutual information $\frac{1}{2}I(X;Y_2|Y_1)$ characterizes the additional information obtained by an extra observation $Y_2=|X|+Z_2$ (due to improved receivers).
Now imagine that we have a different extra observation as $Y_2'=X+Z_2$, then it would lead to a gain of $3$ dB in electrical SNR.
As the SNR increases, the uncertainty of the sign of $X$ vanishes after observing $Y_1$.
Therefore, in the high-SNR limit, the gain from observing $Y_2=|X|+Z_2$ tends to that from observing $\mathrm{sgn}(X)\cdot |X|+Z_2=X+Z_2=Y_2'$.
Thus the high-SNR limit of the gain should be $3$ dB in electrical SNR.

\section{Anaysis of Gap to Capacity}

We have shown that the information rate of ACO-OFDM can be considerably boosted by improved receivers.
However, there is still a significant gap between the information rate $\mathcal I_\textrm G$ and the channel capacity $\mathcal C$ given in (\ref{C}), and the gap is unbounded at high SNR because ACO-OFDM utilizes only a half of degrees-of-freedom (DoF) of the channel.
In this section we analyze this performance loss to gain insight into optical wireless system design.
Our main observation is that, in different SNR regimes, the major cause of performance loss is different.

\begin{figure}
\centering
\includegraphics[scale=0.59]{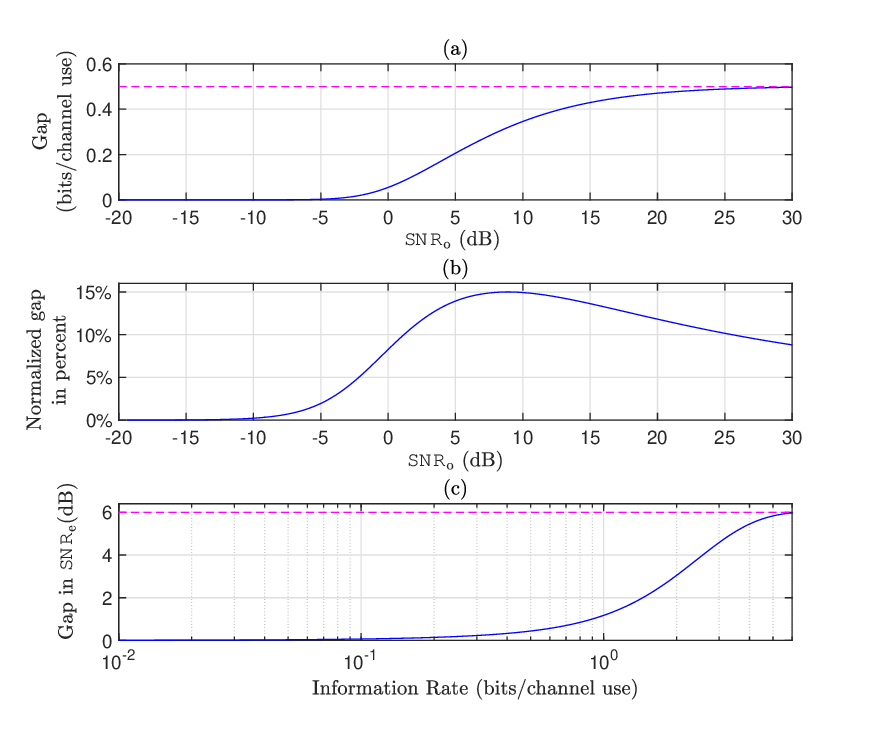}
\caption{Gap between $I(S_\textrm{clip};S_\textrm{clip}+W)$ and $\mathcal I_\textrm{G}$.}
\label{gap1}
\end{figure}

We first consider the loss caused by the dependence between components of the time-domain ACO-OFDM signal $\mathbf S_\textrm{ACO}$.
Since the optical intensity channel (\ref{DTOIC}) is memoryless, to achieve capacity we should let its inputs $\{S_i\}$ be IID \cite{Gallager}.
However, with IID inputs in the frequency domain,\footnote{When capacity-achieving coding over subcarriers is used, the frequency-domain symbols tend to be IID as the code length grows without bound \cite{FWC}.}
the components of $\mathbf S_\textrm{ACO}$ are generally not IID but correlated.
The loss caused by non-IID time-domain inputs can be evaluated by the gap between $\mathcal I_\textrm{G}$, the information rate of ACO-OFDM, and $I(S_\textrm{clip};S_\textrm{clip}+W)$, the information rate of an IID pulse amplitude modulation (PAM) with clipped Gaussian symbols distributed as (\ref{S_f}) (the same as a time-domain sample of ACO-OFDM), where $W\sim\mathcal N(0,\sigma^2)$.
In Figure \ref{Ifigure}, it is shown that the gap between $\mathcal I_\textrm{G}$ and  $I(S_\textrm{clip};S_\textrm{clip}+W)$ is negligible at low SNR, but it becomes considerable at moderate-to-high SNR. 
In particular, we have the following result, which validates the rate loss due to correlated inputs, and shows that the gap is asymptotically $1/2$ bits, or $3$ dB in optical SNR, or $6$ dB in electrical SNR; see Figure \ref{gap1} for related numerical results.

\emph{Proposition 3: The information rate} $\mathcal I_\textrm{G}$ \emph{is upper bounded by}
\begin{align}
\mathcal I_\textrm{G}\leq I(S_\textrm{clip};S_\textrm{clip}+W)
\doteq \frac{1}{4}\log\frac{8\pi\mathcal E^2}{\sigma^2}\doteq\mathcal I_\textrm{G}+\frac{1}{2}\;\textrm{bits/c.u.}\label{clipG}
\end{align}
\begin{IEEEproof}
In Appendix A.
\end{IEEEproof}


\begin{table*}
\centering
\caption{Analysis of gap to capacity for ACO-OFDM with conventional receiver.}
\label{tablegap}
\scalebox{1.1}{
\begin{tabular}{|c|c|c|c|c|}
\hline
\multicolumn{2}{|c|}{\multirow{3}*{~}} & \multicolumn{3}{c|}{\textbf{Magnitudes of Performance Loss}}
\\\cline{3-5}
\multicolumn{2}{|c|}{~}&{\multirow{2}*{\makecell[cc]{\textbf{Caused by}\\\textbf{Suboptimal Receiver}}}}&{\multirow{2}*{\makecell[cc]{\textbf{Caused by}\\\textbf{Correlated Inputs}}}}&{\multirow{2}*{\makecell[cc]{\textbf{Caused by}\\\textbf{Suboptimal Input Distribution}}}}
\\
\multicolumn{2}{|c|}{~}&~&~&~
\\\hline
\multirow{3}*{\makecell[cc]{\textbf{SNR}\\ \textbf{Regimes}}}&\textbf{High}&Asymptotically $1.5$ dB&Asymptotically $3$ dB& \makecell[cc]{Unbounded, asymptotically a \\half of capacity}
\\  \cline{2-5}
~&\textbf{Moderate}&\makecell[cc]{Decreases monotonically as\\SNR decreases (see Fig. 8)} &\makecell[cc]{Decreases monotonically as\\SNR decreases (see Fig. 9)}&\makecell[cc]{Small when $\mathsf{SNR}_\textrm o$ is $0$-$10$ dB}
\\ \cline{2-5}
~&\textbf{Low}&Asymptotically $0.67$ dB&Asymptotically zero &\makecell[cc] {A majority of capacity}
\\ \hline
\end{tabular}}\\
\end{table*}

The remaining gap to capacity, $\mathcal C-I(S_\textrm{clip};S_\textrm{clip}+W)$, can be seen as a performance loss caused by a suboptimal input distribution as (\ref{S_f}).
Numerical results in Figure \ref{Ifigure} show that this gap dominates the performance loss at both high- and low-SNR.
Under an average intensity constraint (\ref{E}), it has been shown that an optimized geometrically distributed input approaches the capacity of (\ref{DTOIC}) to within a few tenths of bits/c.u. \cite{farid2010capacity}.
Specifically, at low SNR the best asymptotic result is obtained by an on-off keying (OOK) input with a high probability of ``off'' (i.e., $S=0$), while at high SNR an exponential input with a zero probability of ``off'' is asymptotically optimal.
The probability of ``off'' of the input distribution (\ref{S_f}), however, is $\Pr\{S=0\}=\frac{1}{2}$,
leading to a waste of DoF at high SNR.
Consequently, in the high-SNR limit, all of $\underline{\mathcal I}_\textrm {G}$, $\mathcal I_\textrm {G}$, and $I(S_\textrm{clip};S_\textrm{clip}+W)$ only achieve a half of the capacity.
However, the gap $\mathcal C-I(S_\textrm{clip};S_\textrm{clip}+W)$ is relatively small at moderate SNR, and it almost vanishes when $\mathsf{SNR}_\textrm{o}$ is around $4$-$5$ dB.

We summarize our analysis of the gap to capacity for ACO-OFDM in Table \ref{tablegap}.
At moderate SNR, ACO-OFDM performs well, and improved receivers play an important role in achieving channel capacity.
In other cases the performance can be enhanced by other methods.
At high SNR, layered optical OFDM schemes can be employed to approach the channel capacity.
A result in \cite{zhou2017capacity} shows that, as the SNR and the number of layers tend to infinity, layered ACO-OFDM with IID complex Gaussian inputs in each layer asymptotically achieves an information rate of $\frac{1}{2}\log(\frac{\pi\mathcal E^2}{8\sigma^2})$, which is within $0.07$ dB to the high-SNR channel capacity.
The result is obtained by employing a conventional receiver in the $\ell$-th layer after an interference cancellation procedure to eliminate the interference from the $(\ell-1)$-th layer if $\ell>1$, combined with a power allocation strategy among layers.
An intuitive interpretation of this near-capacity performance is that the superposition of layers guarantees a vanishing probability of ``off'' and reduces the dependence in the time-domain signal, thereby guaranteeing an optimal pre-log of $1/2$, while a carefully chosen power allocation among layers may achieve a near-optimal input distribution, thereby achieving a small asymptotic SNR gap.
With respect to a given SNR, one may optimize the number of layers and the power allocation to approach the capacity; see, e.g., \cite{YSG17,MCEA19,ZCH20}.
Unfortunately, if the SNR is low, the performance of ACO-OFDM is unsatisfactory even with improved receivers, and no better optical OFDM schemes is known (adding layers is no longer beneficial at low SNR \cite{zhou2017capacity}).
Thus, a simple strategy for system design is to switch to OOK signaling when SNR is below a given threshold (e.g., $0$ dB).

\emph{Remark}:
One may ask whether the performance of layered schemes can be enhanced by replacing the conventional receiver with an improved receiver for each layer.
Although an analytical answer is yet to be found, we expect that the gain of an improved receiver, if any, will be much smaller than that given in this paper.
The reason is that, in this case, the subcarriers that can be utilized by an improved receiver for the $\ell$-th layer are occupied by the unknown data from layers with indices larger than $\ell$, which severely limit the gain.

\section{Extensions and Discussions}

\subsection{Information Rates of ACO-OFDM With Improved Receivers and Practical Constellations}

In this section we still assume the inputs ${\tilde {{X}}_1}$, ${\tilde {{X}}_3},...,{\tilde {{X}}_{N/2-1}}$ to be IID with zero mean and variance $\tilde \sigma_x^2$.
But we alternatively consider alphabets of practical interest, in particular two-dimensional constellations with equiprobable points such as QAM or PSK.
In this setting it is difficult to give an exact characterization like Proposition 1.
Instead, we will establish an upper bound on the information rate of ACO-OFDM in the limit of large $N$.
Numerical results in Section VI will demonstrate that the bound is also relevant for practical values of $N$.

In the limit of large $N$, the central limit theorem guarantees that each component of $\mathbf {X}$ converges in distribution as
$X_i\xrightarrow{\textrm d}\mathcal N(0,\sigma_x^2)$.
Consequently, the time-domain samples of ACO-OFDM have an asymptotic distribution given by (\ref{S_f}).
We then obtain the following approximation. 
\begin{align}
&\mathrm E[S]\approx\frac{\sigma_x}{\sqrt{2\pi}}=\frac{\tilde{\sigma}_x}{2\sqrt{\pi}},\label{approx1}\\
&\mathsf{SNR}_\textrm e\approx\pi\mathsf{SNR}_\textrm o^2.\label{approx2}
\end{align}
However, it should be noted that, in the limit of large $N$,
the asymptotic distribution of the components of $\mathbf {X}$ is not jointly Gaussian unless the symbols ${\tilde {{X}}_1}$, ${\tilde {{X}}_3},...,{\tilde {{X}}_{N/2-1}}$ are Gaussian or complex Gaussian [\ref{Lapidoth}, Theorem 23.6.17].
Consequently, although each pair of components of $\mathbf X$ is asymptotically uncorrelated (i.e., $\mathrm E[X_i X_j]$ converges to zero whenever $i\neq j$), the components as a whole are not asymptotically IID.
Thus the equations (\ref{memoryless1}) and (\ref{memoryless2}) in Section \ref{SecIII}
cannot be used as approximations here,\footnote{For example, for a QAM constellation, it is evident that $\frac{N}{2} I(X;Y_1)$ is bounded away from $I(\mathbf X;\mathbf Y_1)=I(\tilde{\mathbf X}_\textrm{ACO};\tilde{\mathbf Y}_\textrm{odd})=\frac{N}{4}I(\tilde{X};\frac{1}{2}\tilde X+\tilde W)$.}
and we have to begin with (\ref{IACO}) in the general setting.
We obtain the following result which shows that, in practical systems using QAM/PSK constellation, the maximum achievable rate gain of improved receivers is essentially upper bounded by $\Delta$ given in Proposition 1.

\emph{Proposition 4: Under independent and uniformly distributed (IUD) QAM or PSK inputs, the information rate of ACO-OFDM in the optical intensity channel (\ref{DTOIC}) satisfies}
\begin{align}
\lim\limits_{N\to\infty}\mathcal I&\leq\underline{\mathcal I}+\Delta\label{UB},
\end{align}
\emph{where} $\Delta$ \emph{is given in (\ref{singleletter}), and} $\underline{\mathcal I}$ \emph{is the information rate of ACO-OFDM with conventional receiver given in (\ref{I_}), which satisfies}
\begin{align}
\lim\limits_{N\to\infty}\underline{\mathcal I}(\mathsf{SNR}_\textrm o)&<\underline{\mathcal I}_\textrm G(\mathsf{SNR}_\textrm o),\label{UBo}\\
\underline{\mathcal I}(\mathsf{SNR}_\textrm e)&<\underline{\mathcal I}_\textrm G(\mathsf{SNR}_\textrm e),\label{UBe}
\end{align}
\emph{where} $\underline{\mathcal I}_\textrm G$ \emph{is given in (\ref{piE}).}
\begin{IEEEproof}
We first note that
\begin{subequations}\label{IQ}
\begin{align}
\mspace{-12mu}N\cdot\mathcal I&=I(\mathbf {\tilde X}_\textrm {ACO};\mathbf{R}_\textrm {ACO})\\
&=I(\mathbf { X}; \mathbf {Y}_1, \mathbf {Y}_2)\label{IQ3}\\
&=I(\mathbf { X}; \mathbf {Y}_1)+I(\mathbf {X}; \mathbf {Y}_2 | \mathbf {Y}_1)\label{IQ4}\\
&=I(\mathbf {\tilde X}_\textrm {ACO}; \mathbf {\tilde Y}_\textrm {odd})+h(\mathbf {Y}_2 | \mathbf {Y}_1)-h(\mathbf {Y}_2 | \mathbf {Y}_1,\mathbf {X})\label{IQ5}\\
&=\frac{N}{4}I\left(\tilde X; \frac{1}{2}\tilde X+\tilde W\right)+h(\mathbf {Y}_2 | \mathbf {Y}_1) - h(\mathbf {Z}_2)\\
&=\frac{N}{4}I\left(\tilde X; \frac{1}{2}\tilde X+\tilde W\right)+h(\mathbf {Y}_2 | \mathbf {Y}_1) - \frac{N}{2}h(Z_2).
\end{align}
\end{subequations}
By the chain rule for differential entropy and the fact that conditioning reduces differential entropy, we have
\begin{subequations}
\label{dependence}
\begin{align}
&h(\mathbf { Y}_2 | \mathbf { Y}_1)\\
&=h(Y_{2,1} | \mathbf { Y}_1)+\sum\limits_{i=2}^{N/2}h\left(Y_{2,i}|Y_{2,1},...,Y_{2,i-1}, \mathbf { Y}_1\right)\label{d1}\\
&\leq \sum\limits_{i=1}^{N/2}h\left(Y_{2,i}| \mathbf { Y}_1\right)\label{d2}\\
&\leq\sum\limits_{i=1}^{N/2}h\left(Y_{2,i}|  Y_{1,i}\right)\\
&=\frac{N}{2}h\left(Y_{2}|  Y_{1}\right);
\end{align}
\end{subequations}
that is, the dependence among components reduces differential entropy.
Combining (\ref{IQ}) and (\ref{dependence}) yields
\begin{align}
\mathcal I\leq\underline{\mathcal I}+I(X;Y_2 | Y_1),
\end{align}
where $X$ is asymptotically Gaussian in the limit of large $N$, and is exactly Gaussian if and only if
the frequency-domain inputs are complex Gaussian or Gaussian.
Therefore, in the limit of large $N$, $I(X;Y_2 |Y_1)$ converges to $\Delta$, yielding the asymptotic result (\ref{UB}).

To prove (\ref{UBo}) and (\ref{UBe}), we first note that, under IUD QAM or PSK inputs with a given variance $\sigma_{\tilde x}^2$, the information rate $\underline {\mathcal I}$ is upper bounded away from $\underline {\mathcal I}_\textrm G$, which is achievable if and only if the frequency-domain inputs are complex Gaussian.
In the limit of large $N$, the approximation (\ref{approx1}) becomes exact, and an optical SNR constraint on the channel input $S$ is equivalent to a constraint on the variance of $\tilde X$, so that (\ref{UBo}) is obtained.
Since we consider QAM/PSK constellation, the real and imaginary parts of $\tilde X_n$ are symmetric with respect to the origin, respectively.
Then it is straightforward to show that the PDF of $X$ is symmetric with respect to the origin as well.
This symmetry guarantees that $\mathrm E[X|X|]=0$, which implies that $\mathrm E[S^2]=\frac{\sigma_x^2}{2}=\frac{\tilde{\sigma}_x^2}{4}$. 
Thus, an electrical SNR constraint on the channel input $S$ is equivalent to a constraint on the variance of $\tilde X$, so that $\underline {\mathcal I}(\mathsf{SNR}_\textrm e)$ is upper bounded by $\underline {\mathcal I}_\textrm G(\mathsf{SNR}_\textrm e)$ for an arbitrary $N$.
\end{IEEEproof}

In Proposition 4, only an upper bound on $\mathcal I$ is established in a restricted form.
In Section VI we will show that the upper bound can be approached to within a small gap by practical coded modulation schemes.
Then it is reasonable to employ $\underline{\mathcal I}+\Delta$ as a rough approximation of $\mathcal I$ for typical values of SNR and $N$.
However, it should be noted that the optimal distribution of the IID inputs $\{\tilde{X}_n\}$ for ACO-OFDM is still unknown in general.

\subsection{Remarks on Genie Receiver}
The genie receiver perfectly knows the components in $\mathbf R_\textrm{ACO}$ that satisfy $R_i=W_i$ (i.e., $S_i=0$).
We have the following result about its error performance and information rate.

\emph{Proposition 5: The error performance of the genie receiver has a gain of at least} $3$ \emph{dB in electrical SNR or $1.5$ dB in optical SNR over the conventional receiver.
The information rate of the genie receiver, denoted by} $\overline{{\mathcal I}}$, \emph{is lower bounded by }
\begin{align}
\overline{{\mathcal I}}(\mathsf{SNR}_\textrm o)&\ge\underline{\mathcal I}(\sqrt{2}\cdot\mathsf{SNR}_\textrm o),\\
\overline{{\mathcal I}}(\mathsf{SNR}_\textrm e)&\ge\underline{\mathcal I}(2\cdot\mathsf{SNR}_\textrm e),
\end{align}
\emph{and its gain is lower bounded by} $1/4$ \emph{bits/c.u.}

\begin{IEEEproof}
We utilize the additional genie-aided information of genie receiver as follows.
For $0\leq i\leq N/2-1$ we let
\begin{eqnarray}
\label{genie}
{Y_i^\textrm {genie}}  =\left\{ \begin{gathered}
  R_i, \;\;\;\;\;\;\;\;\;\textrm{if} \; S_i\neq 0 \hfill \\
  -R_{i+\frac{N}{2}}, \; \textrm{if} \; S_i=0. \hfill \\
\end{gathered}  \right.
\end{eqnarray}
Then it is straightforward to show that
\begin{equation}
\label{Ygenie}
\mathbf Y^\textrm {genie}=\mathbf X+\mathbf W',
\end{equation}
where $\mathbf Y^\textrm {genie}=[Y_0^\textrm {genie},...,Y_{N/2-1}^\textrm {genie}]$, $\mathbf W'$ is of length $N/2$ and its components are IID $\mathcal N(0,\sigma^2)$ noise samples.
Taking the DFT of
\begin{equation}
[\mathbf Y^\textrm {genie},-\mathbf Y^\textrm {genie}]=[\mathbf X,-\mathbf X]+[\mathbf W',-\mathbf W']
\end{equation}
and noting that a time-domain signal $[\mathbf A,-\mathbf A]$ is a superposition of only the odd subcarriers,
we obtain
\begin{equation}
\label{genieodd}
\tilde Y_n=\tilde X_n+\tilde W'_n,\mspace{4mu}n=1,3,...,N/2-1,
\end{equation}
where $\tilde W'_n\sim\mathcal N(0,2\sigma^2)$, and $\tilde Y_n=0$ if $n$ is even.
By comparing (\ref{genieodd}) with (\ref{FDC}), it is clear that the variance of the noise is reduced by a half, and its standard deviation is reduced by a factor of $1/\sqrt{2}$.
The reduction of the variance of noise guarantees that, compared to the conventional receiver, the genie receiver has a $3$ dB gain in electrical SNR or $1.5$ dB gain in optical SNR in both the error performance and the information rate.
\end{IEEEproof}

Regardless of SNR or input distribution, the genie receiver in the proof of Proposition 5 shows a fixed gain in SNR over the conventional receiver, which is equal to the high-SNR limit of the maximum achievable gain of improved receivers.
According to Propositions 1, 4, and 5, we immediately obtain the following corollary.

\emph{Corollary 1: The genie receiver provides an upper bound on the information rate of ACO-OFDM with IUD inputs in the limit of large} $N$\emph{:}
\begin{equation}
\lim\limits_{N\to\infty}{\mathcal I} \leq\underline{\mathcal I}+\Delta \leq \overline{{\mathcal I}}.
\end{equation}
\emph{For IID complex Gaussian inputs, we have}
\begin{equation}
{\mathcal I}_\textrm G=\underline{\mathcal I}_\textrm G+\Delta \leq \overline{{\mathcal I}}_\textrm G.
\end{equation}

The uncoded BER of the genie receiver was often used as a lower bound of uncoded BERs of improved receivers
(although this bound has never been validated by a proof), and the results in \cite{chen2014improved} as well as our results in Section II-B show that the gap between error performances of improved receivers and the genie receiver decreases as the SNR and the alphabet size increase.
Interestingly, Proposition 5 and Corollary 1 imply that there is a similar relationship in information rate: with respect to the conventional receiver,
the gain of the genie receiver is an upper bound on the maximum achievable gain of improved receivers,
and the gap is relatively small at high SNR.

In fact, the performance gain achieved in the proof of Proposition 5 can be improved.
In both \cite{asadzadeh2011receiver} and our results in Section \ref{IIB}, the genie receiver has a gain of up to $2$ dB in optical SNR, which is considerably larger than the gain observed in \cite{chen2014improved}.
The reason is that the performance of the genie receiver can be enhanced by negative clipping before the operation (\ref{genie}).
That is, we let
\begin{eqnarray}
\label{genie+}
{Y_i^\textrm {genie}}  =\left\{ \begin{gathered}
  \max(R_i,0),  \;\;\;\mspace{4mu} \textrm{if} \; S_i\neq 0, \hfill \\
  \max(-R_i,0),  \; \textrm{if} \; S_i=0, \hfill \\
\end{gathered}  \right.
\end{eqnarray}
and then the corresponding performance is even better.
However, the additional gain is not fixed: it decreases as the SNR and the alphabet size increase.

\subsection{Extension to Related Optical Wireless OFDM Schemes}\label{related}
Flip-OFDM and PAM-DMT are optical wireless OFDM schemes closely related to ACO-OFDM.
With conventional receivers, it is well-known that they achieve essentially the same error performance \cite{fernando2012flip,PAM-DMT}.
In fact, they also achieve essentially the same information rate; see \cite{zhou2017capacity}.
In the following we show that the results about improved receivers in this paper can be straightforwardly extended to Flip-OFDM and PAM-DMT, respectively.
We use similar notation to that of ACO-OFDM.

\subsubsection{Flip-OFDM}
The Flip-OFDM allows using of both odd and even subcarriers but doubles the number of channel uses; see \cite{fernando2012flip} for more details.
Let $\tilde{\mathbf X}_\textrm{Flip}$ be a frequency-domain block of Flip-OFDM and $\mathbf X_\textrm{Flip}$ be its IDFT, respectively.
The transmitter of Flip-OFDM transforms $\mathbf X_\textrm{Flip}$ into two consecutive blocks as a frame of channel input, including a positive block $\mathbf S_1$ whose $i$-th component is $\max(X_i,0)$ and a negative block $\mathbf S_2$ whose $i$-th component is $-\min(X_i,0)$.
The conventional receiver of Flip-OFDM observes only $\mathbf R_1 -\mathbf R_2$ whose $i$-th component is $X_i+ W_{1,i}- W_{2,i}$.
To enhance the performance, improved receivers of Flip-OFDM may utilize $\mathbf R_1+\mathbf R_2$ whose $i$-th component is $|X_i|+ W_{1,i}+ W_{2,i}$.
Comparing the above description of Flip-OFDM with that of ACO-OFDM in Section II, it is clear that essentially the same information theoretic results can be straightforwardly established for Flip-OFDM with improved receivers.

\subsubsection{PAM-DMT}
The PAM-DMT performs asymmetric clipping like ACO-OFDM, while allowing using of both odd and even subcarriers but restricting them to sine functions; see \cite{PAM-DMT} for more details.
Let $\tilde{\mathbf X}_\textrm{PAM-DMT}$ be a frequency-domain block of PAM-DMT.
Its IDFT $\mathbf X_\textrm{PAM-DMT}$ consists of sine subcarriers so that its components satisfy $X_i=-X_{N-i}$.
We denote the first half of $\mathbf X_\textrm{PAM-DMT}$ by $\mathbf X=[X_0,...,X_{N/2-1}]$.
The $i$-th component of a time-domain block $\mathbf S_\textrm{PAM-DMT}$ is $S_i=\frac{1}{2}X_i+\frac{1}{2}|X_i|$,
which implies a decomposition of itself such that $\{X_i/2\}$ consists of sine subcarriers (i.e., $\{\textrm{Im}[\tilde{S}_n]\}$) and $\{|X_i|/2\}$ consists of cosine subcarriers (i.e., $\{\textrm{Re}[\tilde{S}_n]\}$ which is generated by asymmetric clipping) that are orthogonal to sine ones, if and only if $\{\textrm{Re}[\tilde{X}_n]\}$ are set to zero.
The conventional receiver of PAM-DMT observes only the sine subcarriers ($\{\textrm{Im}[\tilde{Y}_n]\}$) and discards the cosine ones ($\{\textrm{Re}[\tilde{Y}_n]\}$),
obtaining an equivalent channel $\textrm{Im}[\tilde{Y}_n]=\frac{1}{2}\textrm{Im}[\tilde{X}_n]+\textrm{Im}[\tilde {W}_n]$, $n=1,...,N-1$.
It can be shown that a one-to-one mapping of the received signal $\mathbf {R}_\textrm{PAM-DMT}$ is $\{\mathbf Y_1,\mathbf Y_2\}$, where
\begin{align}
\mathbf Y_1=\mathbf X+\mathbf Z_1, \mspace{4mu}\mathbf Y_2=|\mathbf X|+\mathbf Z_2,
\end{align}
where $\mathbf Y_j=[Y_{j,1},...,Y_{j,N/2}]$, $j\in\{1,2\}$, $Y_{1,i}=R_i-R_{N-i}=X_i+W_i-W_{N-i}$, $Y_{2,i}=R_i+R_{N-i}=|X_i|+W_i+W_{N-i}$.
To enhance the performance, improved receivers of PAM-DMT may utilize both $\mathbf Y_1$ and $\mathbf Y_2$.
Note that $\textrm{Im}[\tilde{\mathbf Y}]\to\mathbf Y_1$ and $\textrm{Re}[\tilde{\mathbf Y}]\to\mathbf Y_2$ are both one-to-one mappings. Thus, the information rate of PAM-DMT with conventional receiver and improved receivers are given by $\mathcal I(\mathbf X;\mathbf Y_1)$ and $\mathcal I(\mathbf X;\mathbf Y_1,\mathbf Y_2)$, respectively.
Comparing the above description of PAM-DMT with that of ACO-OFDM in Section II, it is clear that essentially the same information theoretic results can be established straightforwardly for PAM-DMT with improved receivers.

\section{Coded Performance of Improved Receivers}

\begin{figure}
\centering
\includegraphics[scale=.64]{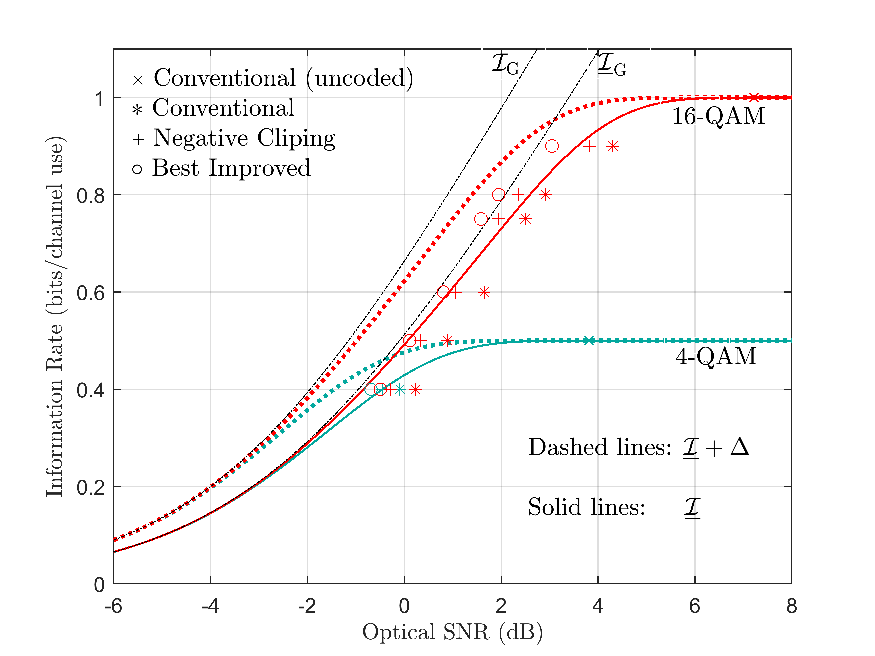}
\caption{Achievable rate of coded ACO-OFDM with improved receivers in optical intensity channel, $\mathsf{BER}=10^{-5}$.}
\label{figure6P8P4P16}
\end{figure}

After establishing information theoretic limits, a question that remains to be answered is how closely can these limits be approached by practical coded modulation schemes.
This section provides numerical performance results of coded ACO-OFDM with different kinds of improved receivers and constellations.
For channel coding we choose the scheme of DVB-S2 \cite{DVB}, which employs an LDPC code concatenated with an outer Bose-Chaudhuri-Hocquenghem (BCH) code and the total block length is $64800$ bits.
The coded bits are mapping to QAM symbols by Gray labeling.
The LDPC decoder performs at most $50$ iterations.
To show that our large-$N$ asymptotic results are relevant for practical values of $N$,
we choose a relatively small DFT/IDFT size of ACO-OFDM as $N=64$.
We consider only the optical SNR in our performance evaluations.
Specifically, we focus on moderate SNR since in that region ACO-OFDM performs better than other schemes; see discussions in the last paragraph of Section IV.
To evaluate the remaining gap to the information theoretic limits, in each figure we show two threshold SNRs corresponding to $\mathcal I$ and $\underline{\mathcal I}+\Delta$, respectively:
(i) The solid vertical line is the minimum SNR required for reliable communication by ACO-OFDM with conventional receiver;
(ii) The dashed vertical line is an approximation of the minimum SNR required for reliable communication by ACO-OFDM with the ML receiver.
Therefore, the gap between them is an approximation of the maximum achievable SNR gain of improved receivers.
The input-constrained information rates $\underline{\mathcal I}$ and the approximations $\underline{\mathcal I}+\Delta$
under different constellations are shown in Figure \ref{figure6P8P4P16}, according to which the two thresholds can be found for a given pair of constellation and code rate.

\begin{figure}
\centering
\includegraphics[scale=0.59]{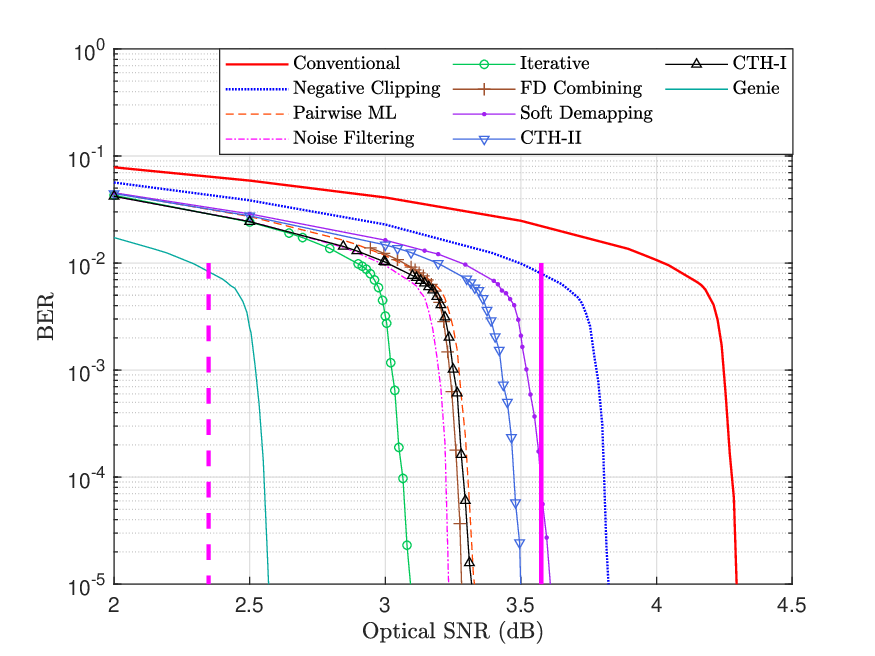}
\caption{BER of LDPC coded ACO-OFDM with improved receivers in optical intensity channel (16-QAM, $\textsf{r}=9/10$).}
\label{figure5M16CR09}
\end{figure}

\begin{figure}
\centering
\includegraphics[scale=0.59]{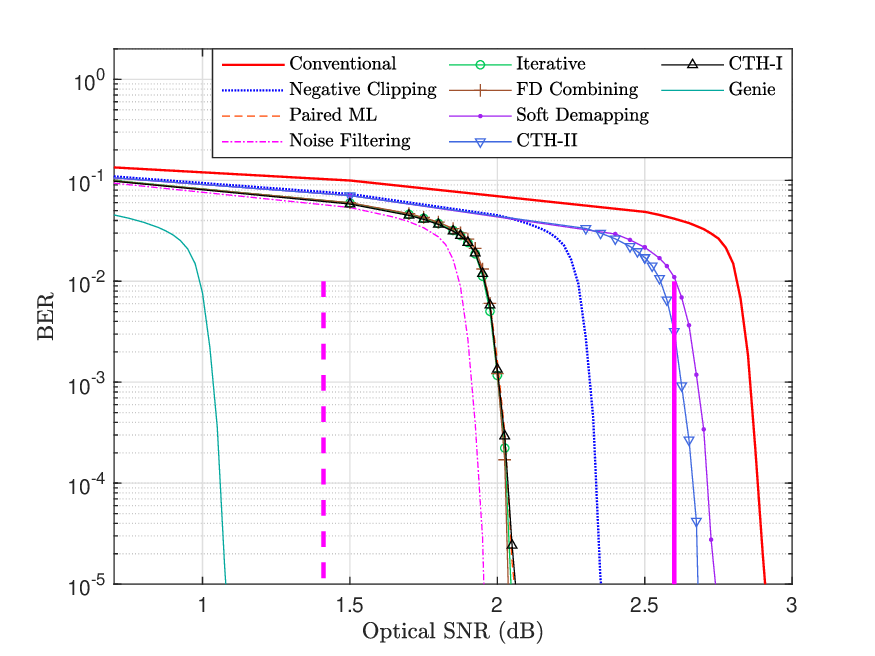}
\caption{BER of LDPC coded ACO-OFDM with improved receivers in optical intensity channel (16-QAM, $\textsf{r}=4/5$).}
\label{figure5M16CR08}
\end{figure}

\begin{figure}
\centering
\includegraphics[scale=0.59]{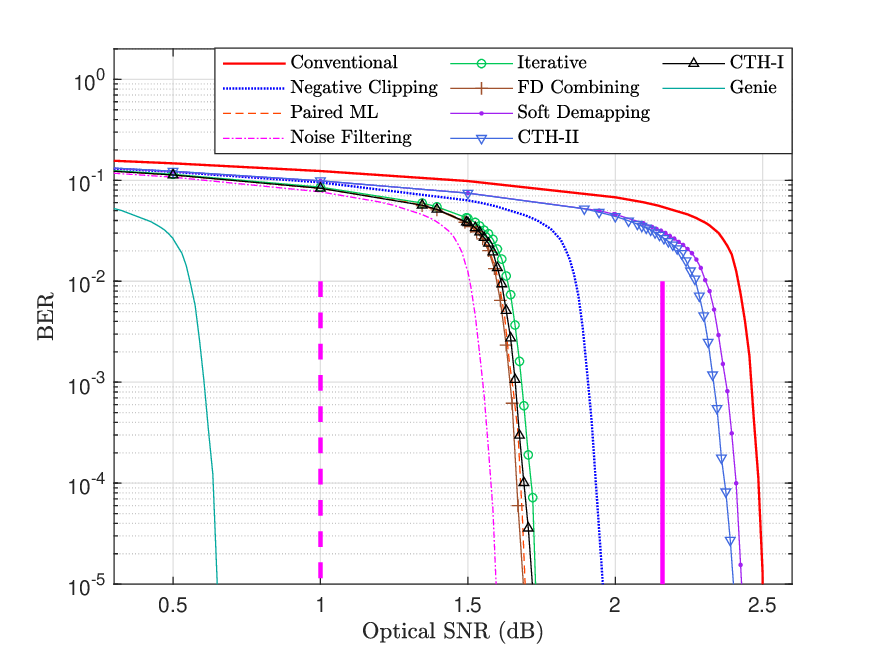}
\caption{BER of LDPC coded ACO-OFDM with improved receivers in optical intensity channel (16-QAM, $\textsf{r}=3/4$).}
\label{figure5M16CR075}
\end{figure}

\begin{figure}
\centering
\includegraphics[scale=0.59]{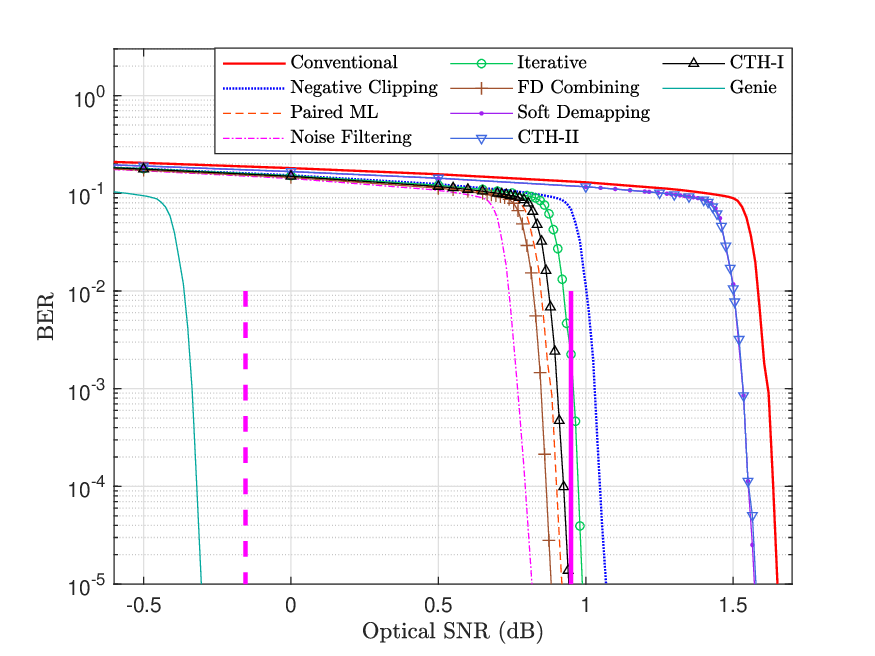}
\caption{BER of LDPC coded ACO-OFDM with improved receivers in optical intensity channel (16-QAM, $\textsf{r}=3/5$).}
\label{figure5M16CR06}
\end{figure}

\begin{figure}
\centering
\includegraphics[scale=0.59]{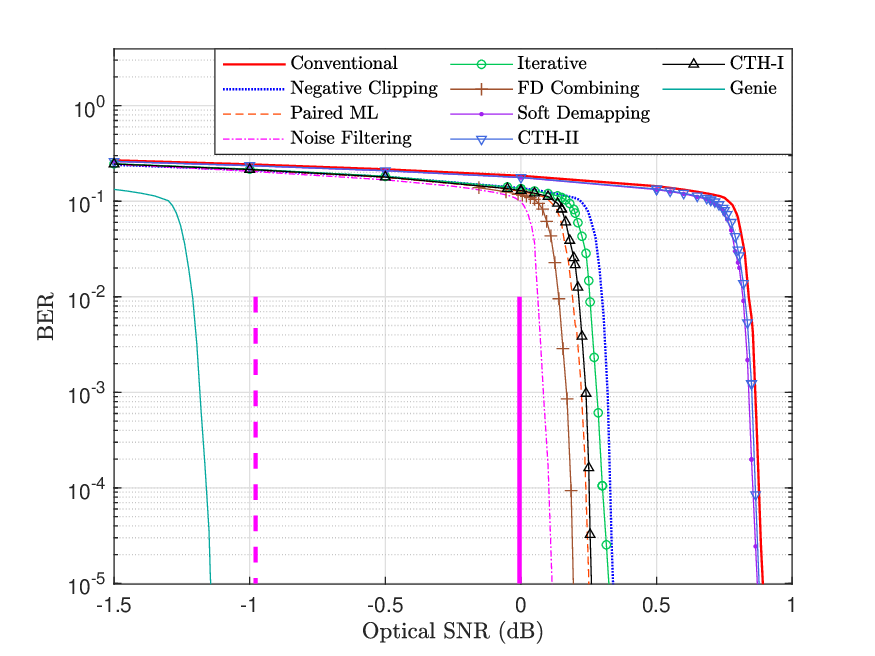}
\caption{BER of LDPC coded ACO-OFDM with improved receivers in optical intensity channel (16-QAM, $\textsf{r}=1/2$).}
\label{figure5M16CR05}
\end{figure}

\begin{figure}
\centering
\includegraphics[scale=0.59]{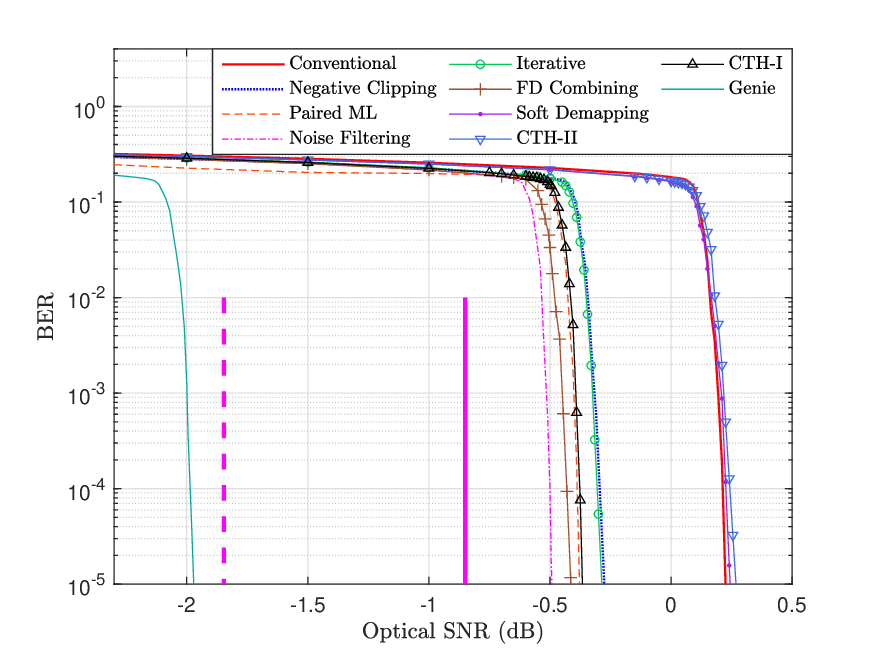}
\caption{BER of LDPC coded ACO-OFDM with improved receivers in optical intensity channel (16-QAM, $\textsf{r}=2/5$).}
\label{figure5M16CR04}
\end{figure}

\begin{figure}
\centering
\includegraphics[scale=0.59]{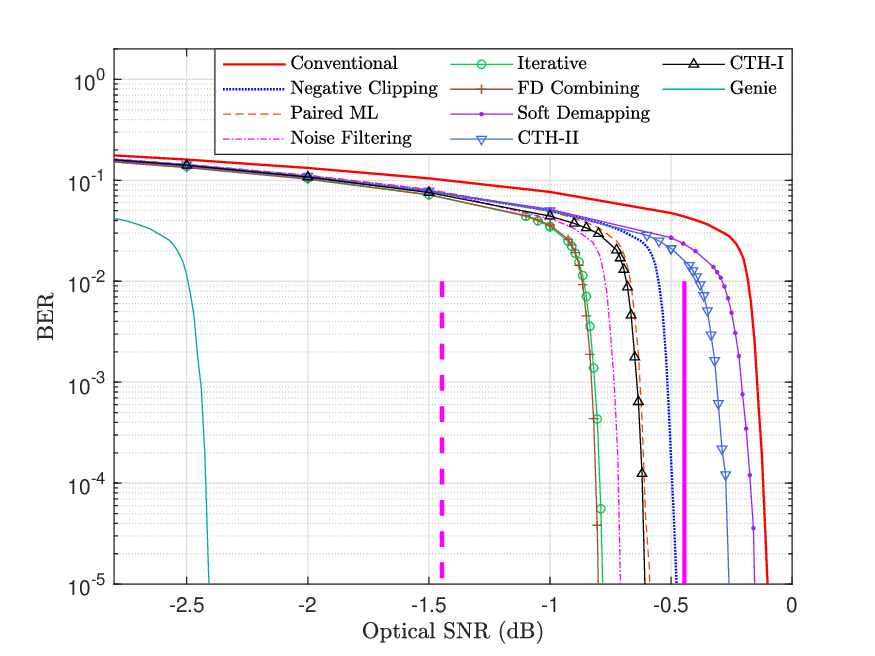}
\caption{BER of LDPC coded ACO-OFDM with improved receivers in optical intensity channel (4-QAM, $\textsf{r}=4/5$).}
\label{figure5M4CR08}
\end{figure}

Our BER performance results are shown in Figures \ref{figure5M16CR09}-\ref{figure5M16CR04}, where we choose some constellations and code rates (denoted by $\textsf r$) for examples.
These error performance results are also summarized in Figure \ref{figure6P8P4P16}.
When the constellation size is $M$, the transmission rate of ACO-OFDM is $\frac{\textsf r \log_2 M}{4}$ bits/c.u.
In the results in Figures \ref{figure5M16CR09}-\ref{figure5M16CR075}, it is shown that the performances of some improved receivers outperform theoretical limits of conventional receiver for high code rates, and the gap to the approximate limit of the ML receiver is within $1$ dB in optical SNR.
For low code rates the gap becomes larger; see Figures \ref{figure5M16CR06}-\ref{figure5M16CR04}.
The performances of improved receivers with 4-QAM in Figure \ref{figure5M4CR08} is better than those with 16-QAM in Figure \ref{figure5M16CR04} at the same total rate ($0.4$ bits/c.u.).
Note that the distance from the threshold SNR of the DVB-S2 code to the Shannon limit in the additive white Gaussian noise channel ranges from $0.7$ to $1.2$ dB in electrical SNR, depending on code rate and constellation size \cite{DVB}.
Thus we conclude that, using typical coded modulation schemes, the performance of ACO-OFDM with existing improved receivers is fairly close to the information theoretic limits we derived, especially at relatively high rates.
Our another observation is that, for improved receivers, a better uncoded performance does not necessarily imply a better coded performance.
For example, the noise filtering receiver \cite{fernando2012flip} performs well in most coded performance examples, although it performs not so well in the uncoded case.

\section{Concluding Remarks}
From an information theoretic perspective, this paper investigates improved receivers of ACO-OFDM and related optical wireless OFDM schemes.
We obtain the following conclusions.
\begin{itemize}
\item
For IID complex Gaussian inputs, the maximum achievable gain of improved receivers is $1/4$ bits per channel use at high SNR, which is equivalent to a gain of $1.5$ dB in optical SNR (or $3$ dB in electrical SNR);
as SNR decreases, the maximum achievable SNR gain of improved receivers decreases monotonically to a non-zero low-SNR limit, which is equivalent to an information rate gain of $36.3\%$.

\item
For constellations such as QAM and PSK, the maximum achievable gain of improved receivers, either in information rate or in optical/electrical SNR, can be roughly approximated by that in the case of IID complex Gaussian inputs.
\end{itemize}
These conclusions imply that improved receivers may reduce the gap to channel capacity significantly at low-to-moderate SNR.
We also give an exact characterization as well as a clear interpretation on the performance of the genie receiver as a performance upper bound for improved receivers.

Finally, we remark that an important future research direction is design and information theoretic limits of improved receivers in optical intensity channels with time dispersion (caused by, e.g., the limited bandwidth of the light-emitting diode).
In the presence of time dispersion, the structure of the ACO-OFDM signal described in Section II-B cannot be utilized directly at the receiver, so that several existing improved receivers are not applicable.

\begin{appendices}
      \section{Proof of Proposition 3}

The inequality in (\ref{clipG}) can be obtained by noting that
\begin{subequations}
\begin{align}
N\cdot\mathcal I_\textrm G&=I(\mathbf {\tilde X}_\textrm {ACO};\mathbf{R}_\textrm{ACO})\notag\\
&=I(\mathbf {S}_\textrm {ACO}; \mathbf{S}_\textrm{ACO}+\mathbf W)\label{Ig1}\\
&=h(\mathbf{S}_\textrm{ACO}+\mathbf W)-h(\mathbf{S}_\textrm{ACO}+\mathbf W|\mathbf {S}_\textrm{ACO})\label{Ig2}\\
&=h(\mathbf{S}_\textrm{ACO}+\mathbf W)-h(\mathbf W)\label{Ig3}\\
&\leq\sum\limits_{i=1}^N \left(h(S_i+W_i)-h(W_i)\right)\label{Ig4}\\
&=N\cdot I(S_{\textrm{clip}};S_{\textrm{clip}}+W),
\end{align}
\end{subequations}
where (\ref{Ig1}) follows because $\mathbf {X}_\textrm {ACO}\to\mathbf{S}_\textrm{ACO}$ is a one-to-one mapping,
(\ref{Ig3}) follows from the independence between $\mathbf W$ and $\mathbf{S}_\textrm{ACO}$, and
(\ref{Ig4}) follows because dependence reduces differential entropy (cf. (\ref{dependence})).

To prove the asymptotics in (\ref{clipG}), we define an indicator random variable as 
\begin{equation}
\label{indicator}
{\mathds{1}}_{S}:=\left\{
\begin{aligned}
1, & & \textrm{if}\;\; S_{\textrm{clip}}> 0&\\
0, & & \textrm{if}\;\; S_{\textrm{clip}}= 0&
\end{aligned} \right.
\end{equation}
so that $S_{\textrm{clip}}={\mathds 1}_{S}\cdot X_\textrm{H}$, where $X_\textrm{H}$ is a random variable following a half-Gaussian distribution with a PDF ${f_{X_\textrm H}}(x) = \frac{2}{{  \sigma _x }}\phi \left( {\frac{{  {x}}}{{\sigma _x}}} \right), \; x\ge 0$.
Then we have
\begin{subequations}
\begin{align}
&I(S_{\textrm{clip}};S_{\textrm{clip}}+W)\\
=\;&I(\mathds 1_S, X_\textrm{H};\mathds 1_S\cdot X_\textrm{H}+W)\\
=\;&I(\mathds 1_S;\mathds 1_S\cdot X_\textrm{H}+W)+I(X_\textrm{H};\mathds 1_S\cdot X_\textrm{H}+W|\mathds 1_S)\\
=\;&h(\mathds 1_S)-h(\mathds 1_S|\mathds 1_S\cdot X_\textrm{H}+W)+\frac{1}{2}I(X_\textrm{H}; X_\textrm{H}+W)\notag\\
&+\frac{1}{2}I(X_\textrm{H};W)\\
=\;&1\; \textrm{bit}-h(\mathds 1_S|\mathds 1_S\cdot X_\textrm{H}+W)+\frac{1}{2}I(X_\textrm{H}; X_\textrm{H}+W).\label{IS}
\end{align}
\end{subequations}

On the other hand, we have
\begin{subequations}\label{hh}
\begin{align}
&I(X;X+W)-I(X_\textrm{H};X_\textrm{H}+W)\\
=\;&h(X+W)-h(W)-\left(h(X_\textrm{H}+W)-h(W)\right)\\
=\;&h(X+W)-h(X_\textrm{H}+W)\\
=\;&h\left(\frac{X+W}{\sigma_x}\right)+\log\frac{1}{\sigma_x}-\left(h\left(\frac{X_\textrm{H}+W}{\sigma_x}\right)+\log\frac{1}{\sigma_x}\right)\\
=\;&h\left(\frac{X}{\sigma_x}+\frac{W}{\sigma_x}\right)-h\left(\frac{X_\textrm{H}}{\sigma_x}+\frac{W}{\sigma_x}\right)\label{hh1}.
\end{align}
\end{subequations}
As the SNR grows without bound, (\ref{hh}) converges to $h\left(\frac{X}{\sigma_x}\right)-h\left(\frac{X_\textrm{H}}{\sigma_x}\right)=h(X)-h(X_\textrm{H})$, which is exactly $1\;  \textrm{bit}$ because $I(X;\mathrm{sgn}(X))=1\;  \textrm{bit}$
and
\begin{subequations}
\begin{align}
I(X;\mathrm{sgn}(X))&=h(X)-h(X|\mathrm{sgn}(X))\\
&=h(X)-\left(\frac{1}{2}h(|X|)+\frac{1}{2}h(-|X|)\right)\\
&=h(X)-h(|X|)\\
&=h(X)-h(X_\textrm{H}).
\end{align}
\end{subequations}
Note that yet another equivalent high-SNR expression of (\ref{IhighSNR}) is given by
\begin{align}
\mathcal{I}_\textrm{G}\doteq\frac{1}{2}I(X;X+W), \label{approx}
\end{align}
where $X\sim\mathcal N(0,\sigma^2)$.
We thus have
\begin{align}
\mathcal{I}_\textrm{G}-\frac{1}{2}I(X_\textrm{H};X_\textrm{H}+W)\doteq\frac{1}{2} \;  \textrm{bits}. \label{approxH}
\end{align}
The proof of the asymptotics in (\ref{clipG}) is completed by comparing (\ref{IS}) and (\ref{approxH}), and noting that the second term of (\ref{IS}) vanishes as the SNR grows without bound.

\end{appendices}

\section*{Acknowledgements}
The authors would like to thank an anonymous reviewer whose comments help us to clarify the square-law relationship between optical SNR and electrical SNR.
Jing Zhou would like to thank Mr. Ru-Han Chen and Prof. Longguang Li for stimulative discussions and helpful comments.

\end{document}